\documentclass[pra,nofootinbib,reprint,graphicx]{revtex4-2}

\usepackage{amsmath}
\usepackage{amsfonts}
\usepackage{array}
\usepackage{color}
\usepackage{comment}
\usepackage{gensymb} 
\usepackage{graphicx}
\usepackage{sidecap}
\usepackage{ulem}
\usepackage{wasysym} 
\usepackage{xspace}

\renewcommand{\em}{\it}

\graphicspath{{../fig/}{./}}

\newcommand{\ie}{{i.e.}\xspace}
\newcommand{\eg}{{\em eg}\xspace}

\newcommand{\rfr}{\ensuremath{_{\text{\tiny ref}}}}

\newcommand{\oned}{\ensuremath{\text{1\sc d}}\xspace}
\newcommand{\threed}{\ensuremath{\text{3\sc d}}\xspace}

\newcommand{\thetak}{\ensuremath{\theta_{\mbox{\tiny k}}}\xspace}

\newcommand{\xref}{\ensuremath{x\rfr}\xspace}

\newcommand{\acronym}[1]{{\text{\sc #1}}\xspace}

\newcommand{\rabbit}{\acronym{rabbit}}

\newcommand{\tdse}{\acronym{tdse}}

\newcommand{\xuv}{\acronym{xuv}}

\newcommand{\ir}{\acronym{ir}}

\newcommand{\scwf}{\ensuremath{\text{\acronym{scwf}}}\xspace}

\newcommand{\HA}[1]{\ensuremath{\mbox{HA}_{#1}}\xspace}
\newcommand{\SB}[1]{\ensuremath{\mbox{SB}_{#1}}\xspace}

\newcommand{\tauxuvir}{\ensuremath{\tau_{\text{\small xuv-ir}}}\xspace}
\newcommand{\wIR}{\ensuremath{\omega_{\text{\sc 0}}}\xspace}
\newcommand{\lIR}{\ensuremath{\lambda_{\text{\sc 0}}}\xspace}

\newcommand{\wigdel}{\ensuremath{\tau_{\mbox{\tiny \sc w}}}\xspace}
\newcommand{\rabdel}{\ensuremath{\tau_{\mbox{\tiny mol}}}\xspace}
\newcommand{\cordel}{\ensuremath{\tau_{\mbox{\tiny probe}}}\xspace}
\newcommand{\avecordel}{\ensuremath{\bar\tau_{\mbox{\tiny probe}}}\xspace}

\newcommand{\Ip}{\ensuremath{E_\acronym{i}}\xspace}
\newcommand{\ket}[1]{\ensuremath{\left\vert#1\right\rangle}}

\newcommand{\Vne}{\ensuremath{V_{\text{\tiny N-e}}}\xspace}

\newcommand{\ener}{\ensuremath{\varepsilon}\xspace}
\newcommand{\der}{\ensuremath{\mbox{d}}}
\newcommand{\kro}[2]{\ensuremath{\delta_{#1#2}}\xspace}
\newcommand{\operator}[1]{\ensuremath{\hat{\mbox{#1}}}}

\definecolor{jcol}{rgb}{1,0,0}
\definecolor{mcol}{rgb}{0.75,0.0,0.75}

\newcommand{\signOK}{\jeremie{\boxed{\mbox{\sc signes ok!}}}}
\renewcommand{\signOK}{}

\definecolor{done}{rgb}{0,0.8,0}
\definecolor{tobedone}{rgb}{0.8,0,0}

\newlength{\figwidth}
\newlength{\figheight}

\begin{document}
\setlength{\figwidth}{0.99\linewidth}
\setlength{\figheight}{0.6\linewidth}

\title{Anisotropic Molecular Photoemission Dynamics Part. I \\ Wigner meets the \rabbit}
\author{Morgan Berkane}
\author{Antoine Desrier}
\author{Camille L\'ev\^eque}
\author{Richard Ta\"\i eb}
\author{J\'er\'emie Caillat}
\email{jeremie.caillat@sorbonne-universite.fr}
\affiliation{Sorbonne Universit\'e, CNRS, Laboratoire de Chimie Physique-Mati\`ere et Rayonnement, LCPMR, F-75005 Paris, France}
\date{\today}

\begin{abstract}
We investigate signatures of anisotropy on the dynamics of time-resolved near-threshold molecular photoemission, through simulations on a one-dimension asymmetric model molecule. More precisely, we study the relationship between the fundamental Wigner delays that fully characterizes the dynamics of one-photon ionization, and the delays inferred from two-color interferometric \rabbit measurements. Our results highlights two different properties pertaining to each of these delays. The first one is related to the inherent necessity to fix an arbitrary electron position origin to define and compute de Wigner-delay. The second one is the dependency of the \rabbit delay on the frequency of the probe laser. Our results show that the angular variations of both delays converge for a specific choice of the position origin and in the limit of a vanishing \ir probe frequency.
\end{abstract} 

\maketitle

\section{Introduction}
The tools of attoscience have allowed revisiting the process of photoemission in the time domain since the late 2000s~\cite{cavalieri2007a,schultze2010a,haessler2009a,klunder2011a}. The ultrafast dynamics revealed by these studies, commonly addressed in terms of scattering time delays~\cite{wigner1955a}, range from few femtoseconds (fs) down to few attoseconds (as). These pioneer experimental studies have been carried out using the  \rabbit\footnote{This acronym
stands for Reconstruction of Attosecond Beatings By Interferences of Two-photon transitions~\cite{muller2002a}.}~\cite{veniard1996a,paul2001a,mairesse2003a} and streaking~\cite{hentschel2001a} schemes, that were initially conceived for the temporal characterization of attosecond light sources~\cite{krausz2009a}. Both can be seen as interferometric pump-probe schemes where a photoemission process is triggered by an attosecond light pulse in the extreme ultra violet (\xuv) regime, and coherently probed by a synchronized infrared (\ir) field. As attosecond science evolved towards atto{\em chemistry}~\cite{salieres2012a,*merritt2021a,*calegari2023a}, the dynamics of photoemission have been investigated in more elaborate molecular systems and nanostructures~\cite{barillot2015a,huppert2016a,bray2018b,ahmadi2022a,boyer2023a,gong2023a,thuppilakkadan2023a},  including chiral species~\cite{beaulieu2017a}. 

This puts forward essential issues regarding an increasing number of degrees of freedom, among which the anisotropy of attosecond-resolved photoemission dynamics~\cite{hockett2016a,zhang2023a,ke2023a}. The latter was first raised in numerical experiments on a model CO molecule where orientation resolved photoemission dynamics were probed using the streaking scheme~\cite{chacon2014a}. These simulations, spanning photoelectron energies up to $\sim 80$ eV, evidenced ``stereo'' delays reaching up to several tens of as. The stereo streaking delays measured in this work accurately match the angular variations of the actual scattering delays over most of the covered energy range. However, discrepancies showing up at lower energies, below $\sim20$ eV, suggest a significant influence of the probe in the measured temporal asymmetries.  Indeed,  \rabbit measurements of photoemission delays in He atoms using \rabbit~\cite{heuser2016a} have highlighted the symmetry breakdown induced by the two-photon probe process itself ~\cite{hockett2017a,bray2018a,fuchs2020a}. Orientation resolved \rabbit experiments further evidenced the imprint of the initial and final states asymmetry on  photoemission dynamics in Ar atoms~\cite{cirelli2018a} and dimers~\cite{trabert2023a}.  The first experiments  reporting anisotropic molecular photoemission dynamics were also performed on CO, using \rabbit in the $0-20$ eV photoelectron energy range~\cite{vos2018a}. The results are consistent with the streaking simulations of~\cite{chacon2014a}, evidencing an anisotropy reaching several tens of as, with `faster' photoemission from either the C or O side of the molecule depending on the photoelectron energy. By comparing photoemission from different electronic channels, this work evidenced experimental imprints of the initial electron position within the molecule in the measured orientation dependent delays~\cite{desrierphdthesis}. In the same spirit, resonant asymmetric photoemission delays were recently investigated experimentally in NO~\cite{gong2022a}, also using \rabbit but with improved angular resolution. These measurements evidence delays varying by few to several tens of as when scanning the photoemission angle in the molecular frame, within the $20-40$ eV energy range.  
In ~\cite{liao2021a}, numerical \rabbit simulations on model molecules similar to the one used in~\cite{chacon2014a} evidenced a probed-induced asymmetry of the measured delays amounting to up to 100 as near ionization threshold, and persisting significantly over several tens of eV. These large values were attributed to the choice of the origin in the definition of the Wigner delays. 

Beyond the fundamental time-domain interpretation of photoemission scattering delays~\cite{schultze2010a,maquet2014a}, the question of their measurement and notably of the influence of the probe in attosecond pump-probe experiments, has been the subject of an important theoretical activity, see \eg~\cite{dahlstrom2012a,dahlstrom2013a,pazourek2015a,cattaneo2016a,vacher2017a,kheifeits2023a}. However, available analytical models to date accurately predict the dependency of the measurements with respect to probe parameters such as the wavelength, but none account explicitly for the anisotropy of the pertubation induced by the measurement.

In this paper, we clarify the role of the arbitrary origin in the definition and computation of Wigner delays, and study the imprint of the average initial electron position in delays measured using the \rabbit technique. This indirectly raises issues regarding the role of the probe wavelength in \rabbit measurements. Our work is based on numerical simulations performed on a \oned asymmetric model molecule. The paper is organized as follows. The model molecule is presented in Sec.~\ref{sec:numtool}. The influence of the electron position origin on the orientation-resolved Wigner scattering delay is investigated in Sec.~\ref{sec:wigner} and the influence of the probe wave-length in orientation-resolved \rabbit measurements independently in  Sec.~\ref{sec:rabbit}. Eventually, we address in Sec.~\ref{sec:tauCC} the anisotropic differences between the fundamental Wigner delays and \rabbit measurements, which inherit the origin-dependency of the former and the wavelength-dependency of the latter. 

By default, equations are expressed in atomic units (a.u.).
\section{Numerical toolbox}\label{sec:numtool}
\subsection{Model molecule}\label{sec:model}
We performed our simulations on a one-dimensional (\oned) model molecule, reminiscent of the ones used {\em eg} in~\cite{chacon2014a,liao2021a}. It is made of a single active electron initially bound to an asymmetric effective potential. Its hamiltonian reads
\begin{eqnarray}\label{eqn:H0}
H_0&=&-\frac{1}{2}\frac{\partial^2}{\partial x^2}+\Vne(x)
\end{eqnarray}
where $x$ is the electron coordinate, and
\begin{eqnarray}\label{eqn:Vne}
\Vne(x)&=&-\frac{q}{\sqrt{(x-X_1)^2+a^2}}-\frac{(1-q)}{\sqrt{(x-X_2)^2+a^2}}.
\end{eqnarray}
 is an asymmetric double soft-core potential. The parameters $X_1,X_2$ represent the (fixed) positions of the nuclei, $q$ is an effective atomic charge and $a$ is an arbitrary screening constant. This set of parameters are numerical knobs that can be tuned to assign the model some desired properties. Here we set the charge $q$ to $0.33$ a.u. and the internuclear distance $R=\vert X_2-X_1 \vert$ to 1.115 \AA. The screening parameter ($a=0.198$ \AA)
  was adjusted to obtain an ionization potential $\Ip=29.77$ eV. The resulting potential function is shown in Fig.~\ref{fig:wfn}(a). It was thus designed with some arbitrariness to obtain realistic molecular features, and more specifically to reflect the difference of electronegativity between the two atoms in an heteronuclear molecule. The asymmetry is  reflected in the ground state wave-function $\Phi_0(x)$, also displayed in Fig.~\ref{fig:wfn}(a), which is the initial state in all the simulations presented hereafter.
 
 \subsection{\oned polar coordinates}
In the present work, we solved the time-independent Schr\"odinger equation (see \eg Sec.~\ref{sec:scws}) using a partial-wave expansion consisting in expressing any function of $x\in\mathbb{R}$ in terms of its odd and even components. It relies on  \oned `polar' coordinates for the electron position, namely a radius $r=\vert z \vert \in \mathbb{R}^+$ and an angle $\theta=\arccos(z/\vert z \vert) \in \{0\degree,180\degree\}$, where $z=x-x\rfr$ is the `cartesian' coordinate $x$ referred to an arbitrary origin $x\rfr$. More details about this approach are provided in Appendix~\ref{sec:pwexpand}.
This technical choice has no practical consequence on the results presented hereafter, which are numerically converged. Nevertheless, it tightens the analogies between the present work and usual approaches invoked in three dimensions (\threed), see \eg~\cite{huppert2016a}. As will be discussed further, it highlights the arbitrariness of the origin of the partial wave expansion,  \ie the origin \xref chosen here to discriminate the right and left sides of the molecule . 

All through the paper, \thetak represents the direction of photoemission, restricted to two discrete values $\thetak=0\degree$ (emission towards the right, $x>0$) and $180\degree$ (towards the left, $x<0$). 

\section{Fundamental dynamics: anisotropic Wigner delays}\label{sec:wigner}
We first studied how the asymmetry of the model molecule is reflected in the so-called stereo Wigner delays~\cite{chacon2014a,vos2018a}, \ie the relative photoelectron scattering delay  towards one side of the molecule (here left) in comparison to the other (right). 

 \begin{figure}[t]
\includegraphics[width=\figwidth]{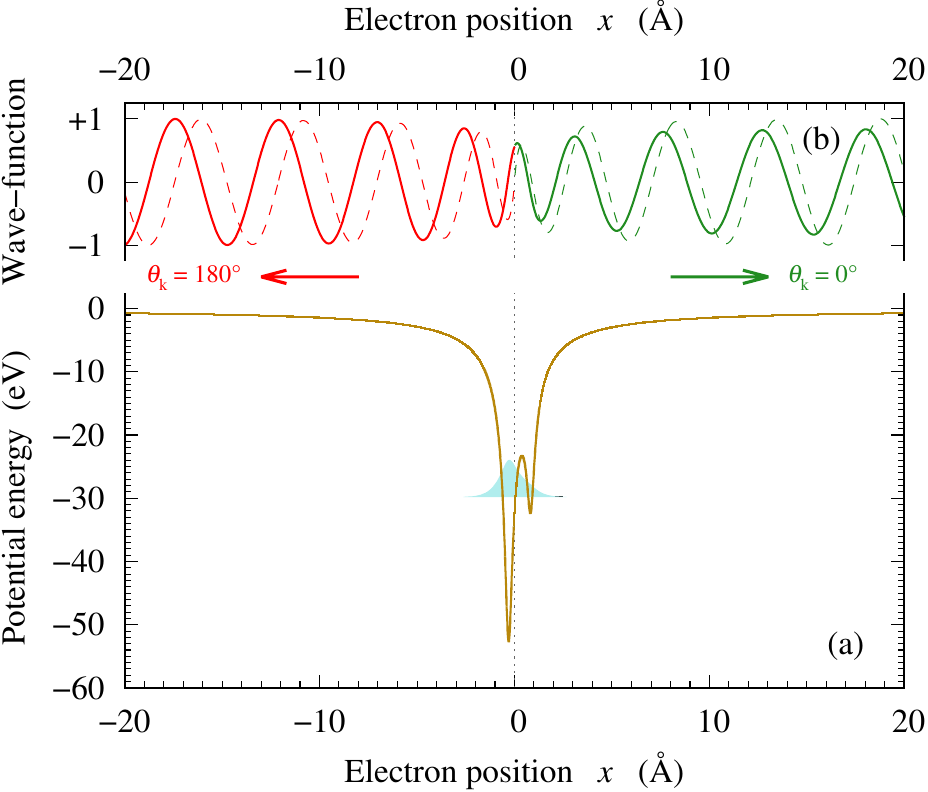}
\caption{\label{fig:wfn} \oned model molecule. (a) Electron-nuclei potential $\Vne(x)$ as a function of the electron position $x$ (dark yellow full curve). The ground state electronic wave-function is also shown (light blue filled curve). (b) Electronic continuum wave-function (full curve) selected at the energy $\varepsilon=4.35$~eV by a one-photon transition from the electronic  ground state. Odd-parity reference wave-function (dashed curve) used to define and compute the orientation-dependent phase-shifts. The displayed continuum wave-functions are normalized such that their amplitudes asymptotically converge to 1 on the left-hand side of the molecule.
In this Figure, the left/right discrimination and the parity refer to the arbitrary $x=0$ position (indicated by a vertical dotted line). }
\end{figure}

\subsection{Selected continuum wave-function}\label{sec:scws}
To this end, we computed and analyzed the continuum-wave functions selected by 1-photon ionization processes (\scwf). The \scwf formalism~\cite{gaillac2016a} is a convenient approach which unambiguously separates the computation of electronic continuum wave-functions in the framework of photoemission from their analysis in terms of scattering dynamics. For a given transition leading to a final energy \ener, the selected continuum state is defined in \oned as~:
\begin{eqnarray}\label{eqn:scwf}
\ket{\Psi_{\varepsilon,\mbox{\tiny sel}}}=\sum\limits_{k=1,2} \langle \Psi_{\ener,k} \vert \operator{d} \vert \Phi_0 \rangle \ket{\Psi_{\ener,k}}
\end{eqnarray}
where \ket{\Phi_0} is the initial bound state, \operator{d} is the dipole operator associated with the process and $\{\ket{\Psi_{\ener,1}},\ket{\Psi_{\ener,2}}\}$ is an {\em arbitrary} orthonormal eigenstate basis for the doubly degenerate continuum at the considered energy. 

As an illustration, the \scwf computed for $\varepsilon=4.35$ eV (full curve)  is shown in Fig.~\ref{fig:wfn}(b). It corresponds to an ionization triggered by a $35.67$ eV photon, \ie the 21st harmonic of a 800 nm \ir laser. Photoemission dynamics are encoded in the spectral variations of the phases of its asymptotic oscillations~\cite{gaillac2016a}. The origin chosen  to discriminate the right and left sides of the molecule is here  the origin of the $x$ axis  (\ie $x\rfr=0$), which is itself arbitrary.
\begin{figure*}[t]
\includegraphics[height=\figheight]{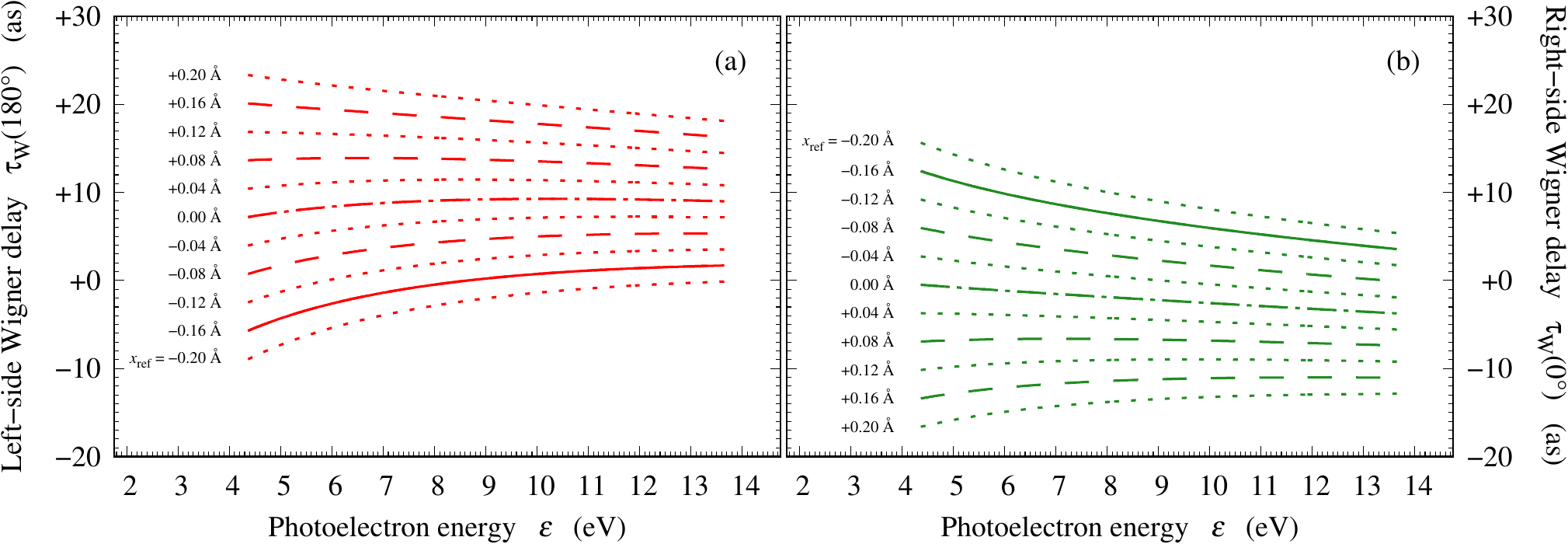}
\caption{\label{fig:wigdel}  Orientation-resolved Wigner delays (Eq.~\ref{eqn:wigdel}) towards the left (a) and the right (b) sides of the molecule as a function of the photoelectron energy. The data are displayed for a set of reference positions $x\rfr$ discriminating the left and right sides of the molecule. The delays are defined with respect to Coulomb waves centered at $x\rfr$. The $x\rfr=0$ results (dashed-dotted curve) were obtained through  \scwf computations and analysis (see text), and were used to infer the $x\rfr\neq0$ results (alternating line styles, see labels) using the $\varepsilon$- and $x\rfr$-dependent correction given in Eq.~\ref{eqn:dwigdel}.  The full curves correspond to the data obtained for $\xref=\langle x \rangle_0$ (see Eq.~\ref{eqn:x0}). \signOK}
\end{figure*}

\subsection{Orientation-resolved Wigner phase-shifts and delays}

The orientation-dependent phase shifts of the \scwf, $\eta(\thetak)$, are defined and computed with respect to an arbitrary intermediate reference wave at the same energy. For the latter, we took the radial $s$ Coulomb wave of the hydrogen atom centered at $x=x\rfr$, displayed as a dashed curve in Fig.~\ref{fig:wfn}(b). This is a typical choice to characterize photoemission dynamics in terms of delays, since molecular continuum waves behave asymptotically as Coulomb waves. Note that these phase shifts depend {\em by definition} on the origin $x\rfr$ chosen to discriminate the left and right sides of the molecule -- or, in other words to set up the \oned partial-wave expansion in terms of even and odd components when computing the \scwf, see Appendix~\ref{sec:pwexpand}.

The orientation-resolved Wigner delays are then defined as the group delay 
\begin{eqnarray}\label{eqn:wigdel}
\wigdel(\thetak)&=&\frac{\partial \eta(\thetak)}{\partial\varepsilon}.
\end{eqnarray}
When computing them, among a series of numerical tests, we ensured that the phase shifts were evaluated in a region where they do not depend on their computation distance on either side of the origin, \ie in the asymptotic region where the potential is symmetric. However, they naturally inherit the $x\rfr$-dependency from $\eta(\thetak)$.

\begin{table}
\begin{tabular}{|rr|rrr|}
\hline
$x\rfr$ & (\AA) & $-0.20$ & $\textcolor{white}{+}0.00$ & $+0.20$ \\
\hline\hline
$\wigdel(0\degree)$ & (as) & $+15.6$ & $-0.5$ & $-16.5$ \\
$\wigdel(180\degree)$ & (as) & $-8.9$ & $+7.2$ & $+23.3$ \\
\hline 
\end{tabular}
\caption{\label{tab:wigdel} Orientation-resolved Wigner delays $\wigdel(\thetak)$ (Eq.~\ref{eqn:wigdel}) computed at $\varepsilon=4.35$ eV for three values of the reference position $x\rfr$ arbitrarily discriminating the right ($\thetak=0\degree$) and left  ($\thetak=180\degree$) sides of the asymmetric model molecule. }
\end{table}
The dependency of the Wigner delays with respect to this analysis parameter is illustrated in Table~\ref{tab:wigdel}, with the values of $\wigdel(\thetak)$  obtained at $\varepsilon=4.35$ eV for $x\rfr=-0.20$, $0$ and $+0.20$ \AA\ respectively. With these data, one can verify that shifting the origin from 0 to a given value $x\rfr$ is equivalent to modifying the path difference between the photoelectron and the arbitrary intermediate reference by $-x\rfr$ on the right side, $+x\rfr$ on the left side. Indeed, it induces a delay shift on each side of the molecule which can be accurately modeled as~\cite{desrierphdthesis,vos2018a} 
\begin{eqnarray}\label{eqn:dwigdel}
\delta\wigdel(\thetak)&=&\cos\thetak \frac{x\rfr}{\sqrt{2\varepsilon}},
\end{eqnarray} 
\ie the time needed for an electron with constant velocity $\sqrt{2\varepsilon}$ to cover the path difference $\delta x = + x\rfr$ towards the left, $-x\rfr$ towards the right.  

\begin{figure}[t]
\includegraphics[height=\figheight]{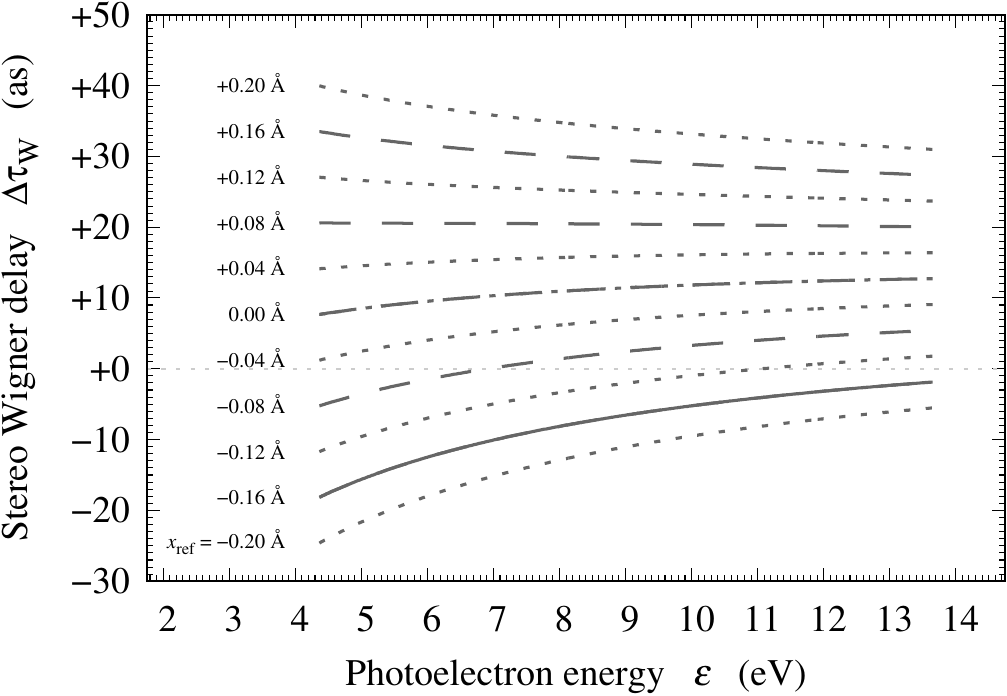}
\caption{\label{fig:wigdel_diff}  Stereo Wigner delays $\Delta\wigdel$ (Eq~\ref{eqn:wigdel_diff}) as a function of the photoelectron energy $\varepsilon$, computed for a set of reference positions $x\rfr$, see caption of Fig.~\ref{fig:wigdel}. \signOK}
\end{figure}

Our simulations covered photoelectron energies ranging over $\sim10$ eV near threshold. The obtained orientation-resolved delays towards the left and right sides of the molecule are displayed in frames (a) and (b) of Fig.~\ref{fig:wigdel} respectively, for a series of $x\rfr$ values comprized between $-0.20$ and $+0.20$ \AA. They  typically lie in the attosecond range. It is noteworthy that such small, sub-\AA, displacements of the origin (which here remain `within' the molecule) induce considerably large relative variations of the delays, with no converging pattern (see the linear dependency given by  Eq.~\ref{eqn:dwigdel}). 

These orientation-resolved delays depend not only on an arbitrary origin, but also on the nature of a reference system (here Coulomb waves) which is arbitrary too.  This last dependancy nevertheless vanishes when considering the stereo-Wigner delays~\cite{chacon2014a,vos2018a,liao2021a}, defined here as the difference
\begin{eqnarray}\label{eqn:wigdel_diff}
\Delta\wigdel=\wigdel(180\degree)-\wigdel(0\degree).
\end{eqnarray}
Figure~\ref{fig:wigdel_diff} shows the values of $\Delta\wigdel$ obtained with the orientation-resolved delays commented above (Fig.~\ref{fig:wigdel}). They are here of the same order of magnitude than the delays $\wigdel(\thetak)$ themselves. We have verified with few test cases that these values remain unchanged when using intermediate reference plane waves instead of Coulomb waves. Yet, the $x\rfr$-dependency remains just as pronounced as for the orientation-resolved delays.\footnote{It is important to note that the origin dependency concerns only the orientation-dependent (or partial-wave) {\em analysis} of the \scwf, but not the \scwf itself which is by essence origin invariant. It therefore impacts the phase shifts and the associated group delays, but not observables such as the orientation dependant ionization {\em probabilites} (which are proportional to the squared asymptotic amplitudes of the \scwf on each side of the molecule~\cite{gaillac2016a}). }

The linear dependency of the stereo Wigner delay with respect to the origin, which is considered to be the main cause for the channel-dependent delay variations reported in~\cite{vos2018a}, is inherent to its very definition. 
In practice, the most natural choice consists in setting the origin to the average electron position in the initial state,
\begin{eqnarray}\label{eqn:x0}
\langle x \rangle_0&=&\int\limits_{-\infty}^{+\infty} x \vert \Psi_0(x) \vert^2\, \der x.
\end{eqnarray}
Characterizing the initial state with this average position is as relevant as characterizing the photoemission dynamics, in the final state, in terms of group delays\footnote{The question of the origin-dependency is  generally overlooked in studies dealing with atoms or centrosymmetric molecules, including in standard textbooks on scattering theory, where it is ``naturally'', implicitly and undisputedly set at the center of symmetry for obvious practical reasons.}. Here, $\langle x\rangle_0=-0.16$ \AA. The corresponding data in Figs.~\ref{fig:wigdel} and~\ref{fig:wigdel_diff} are displayed in full lines.

This arbitrary choice is also comforted by theoretical studies regarding particle scattering in \oned anisotropic potentials~\cite{saalmann2023a}. Nevertheless, it remains to be confronted to the context in which the stereo Wigner delays are investigated experimentally, using interferometric schemes based on the \rabbit~\cite{paul2001a} and streaking~\cite{hentschel2001a} setups. In the present work, we focus on the \rabbit approach, as discussed below.    
\section{\rabbit simulations: anisotropic molecular delays} \label{sec:rabbit}

In this section, we investigate the anisotropic photoemission dynamics inferred from \rabbit measurements. This part is {\em a priori} independent of the previous section since, as we will see, the dependencies of the `\rabbit delays' are fundamentally different from the one discussed for the Wigner delays. 

 \begin{figure}[t]
 \includegraphics[width=\figwidth]{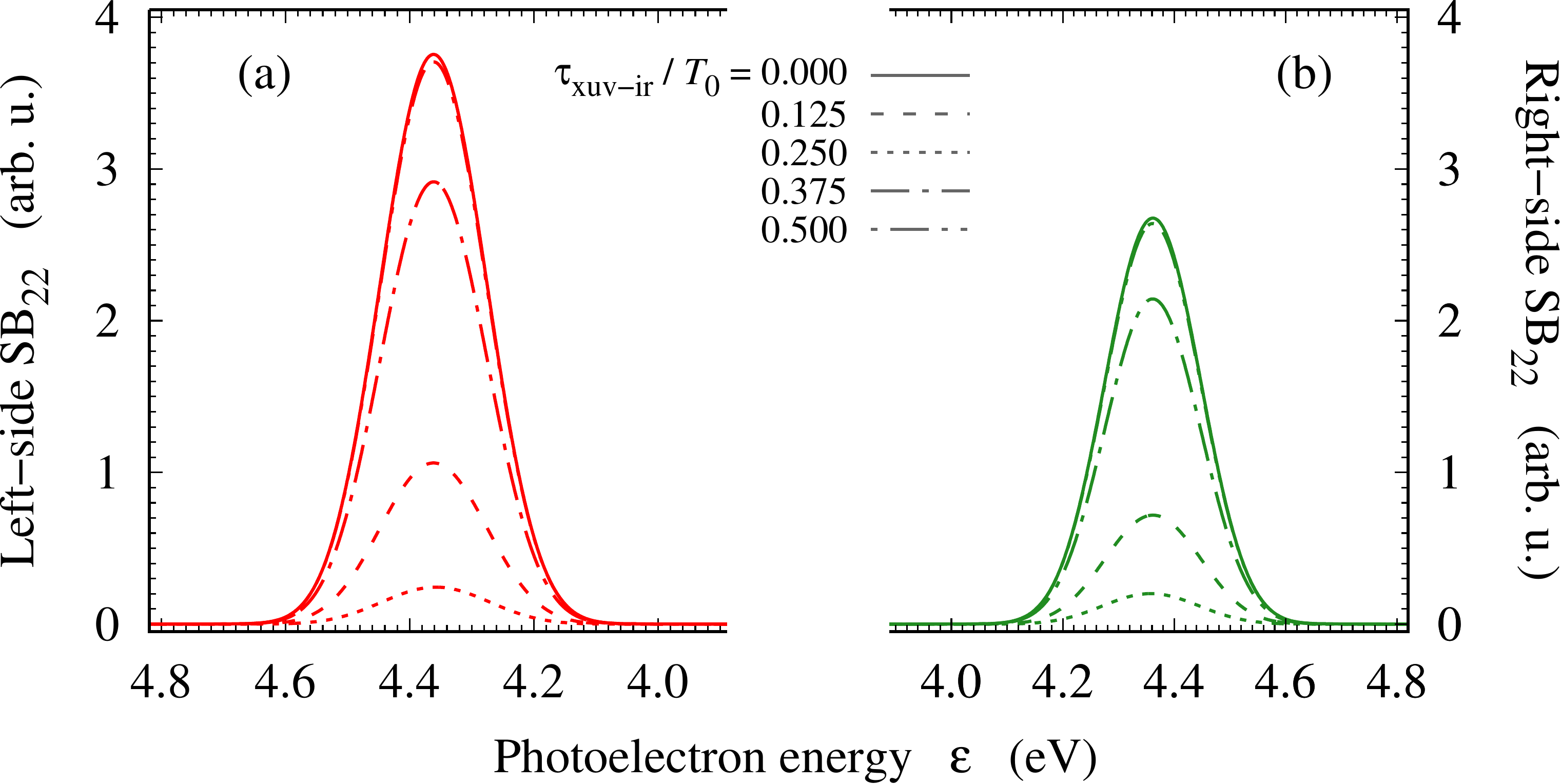}
\caption{\label{fig:rabspc} Orientation-resolved \SB{22} towards the left (a) and right (b) sides of the model molecule, in \rabbit simulations at 800 nm probe wavelength. The spectra are shown for five values of the pump-probe delay \tauxuvir (see inset key) sampling a complete oscillation, see Eq.~\ref{eqn:genrab}.
}
\end{figure}
Following the \rabbit scheme, we simulated photoemission from our asymmetric model molecule with a comb of extreme ultraviolet (\xuv) odd harmonics of an infrared (\ir) laser, dressed by the fundamental field with frequency \wIR. 
We solved numerically the time-dependent Schr\"odinger equation starting from the ground state, the dipole interaction with the \xuv and \ir pulses being implemented in the velocity gauge. The light pulse vector potentials were all assigned $\sin^2$ temporal profiles lasting 40 fs (15 periods of a 800 nm laser, full durations). The harmonics were synchronized with no `attochirp', for the sake of simplicity, but were time-shifted by an adjustable delay \tauxuvir with respect to the \ir  pulse (relative to their maxima). We set the intensities safely in the perturbative regime, notably to prevent any significant transitions involving more than one \ir photon. 
\begin{figure*}[t]
\includegraphics[height=\figheight]{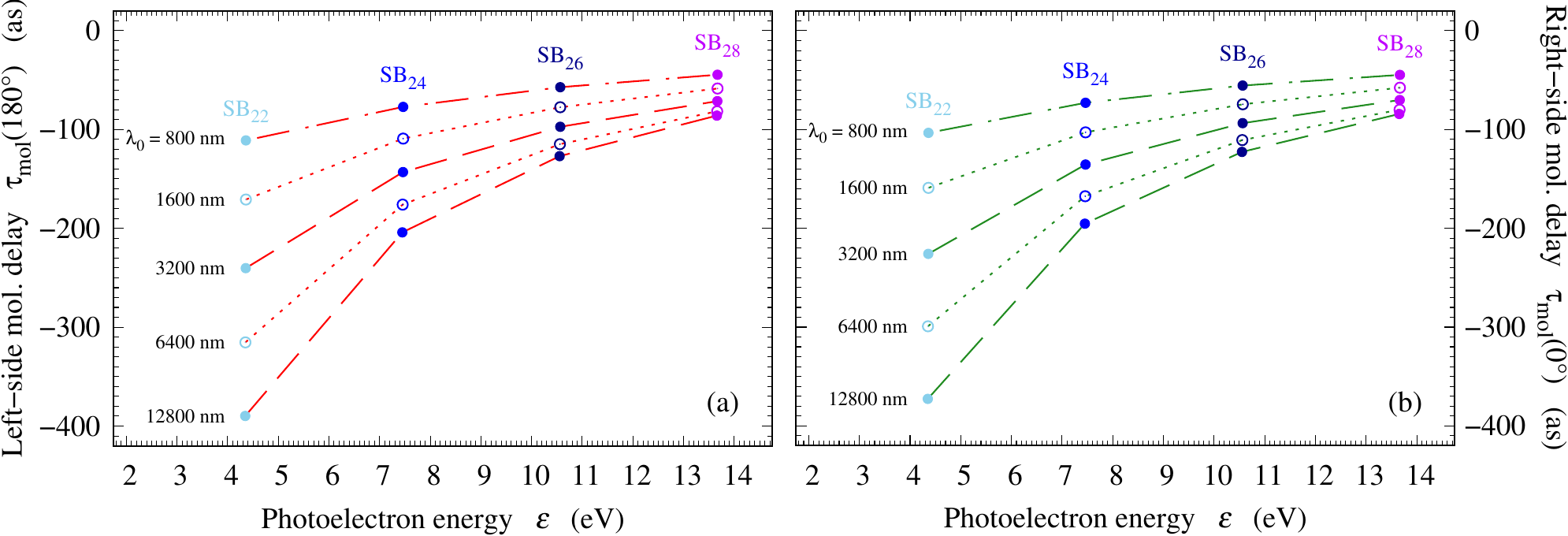}
\caption{\label{fig:rabdel} Orientation-resolved molecular delays (Eq.~\ref{eqn:rabdel}) measured on the left (a) and right (b) sides of the molecule as a function of the average sideband energy. The data were obtained in a series of \rabbit simulations with various `probe' wavelength values (see labels).  The actual data, shown as symbols, are connected with guidelines for each value of \lIR (with alternating line styles). For simplicity, the sideband labels are displayed for the standard $800$ nm case only (dashed-dotted lines).  \signOK
}
\end{figure*}
\subsection{Photoelectron spectra}

Orientation-resolved photoelectron spectra were computed using narrow spectral filters~\cite{kulander1992a} applied to the left and right sides of the wave-function at the end of the simulations. In order to safely discriminate photoelectrons leaving towards each side of the molecule, we let the wave-function propagate few fs after the end of the pulses. In the present context, we emphasize that these spectra do not depend on the origin set to discriminate the right and left sides of the molecule, consistently with their experimental equivalent. 

The spectra consist in a series of main peaks (\HA{2q+1}), evenly separated by a $2\wIR$ energy gap, associated with the absorption of each harmonic. Additional sidebands, each induced by 2-photon transitions involving one harmonic and one \ir photon, show up in between. Each sideband (\SB{2q}) results from two interfering pathways: (i) absorption of both harmonic $2q-1$ and an \ir photon and (ii) absorption of harmonic $2q+1$ and stimulated emission of an \ir photon. Therefore the intensity of each sideband, on each side of the molecule, oscillates at a $2\wIR$ frequency when scanning \tauxuvir~\cite{veniard1996a}. 

As an illustration, Fig.~\ref{fig:rabspc} shows the \SB{22} \rabbit spectra obtained with a standard fundamental photon energy of $\wIR=1.55$ eV, which corresponds to a wavelength $\lIR=2\pi c/\wIR = 800$ nm ($c$ is the light velocity). Frames (a) and (b) display the spectra obtained on the left and right sides of the molecule, respectively, for 5  values of \tauxuvir covering a full oscillation in the \rabbit spectrogram. They reveal a clear asymmetry in the overall ionization probability, in favor of the left side (a similar asymmetry is observed in the 1-photon case, see the \scwf amplitudes on Fig.~\ref{fig:wfn}). They also highlight the typical  \tauxuvir-dependency of the SB magnitudes inherent to \rabbit, on each side of the molecule. 

\subsection{Orientation-resolved molecular phases and delays}
Information on the photoemission dynamics are encoded in the phases of the sideband oscillations,  so-called `atomic' (or, here, `molecular') phases~\cite{veniard1996a}. While the orientation dependency of these oscillations are hardly visible on Fig.~\ref{fig:rabspc}, we will see that they are significant when interpreted in the time domain at the attosecond  scale. 

In our simulations, orientation-resolved molecular phases $\vartheta(\thetak)$ were extracted from the photoelectron spectra by fitting the interferometric pattern function~\cite{veniard1996a}:
\begin{eqnarray}
f(\tauxuvir)&=&P(\thetak)+Q(\thetak)\cos[2\wIR\times
\tauxuvir-\vartheta(\thetak)] \nonumber \\ \label{eqn:genrab}
\end{eqnarray} 
to the spectrally integrated intensity of each sideband on each side of the molecule ($\thetak=0\degree,180\degree$). All the $\thetak$-dependent quantities play the role of fitting parameters.

Various schemes have been developed over the past years to analyze these phases in the time domain, see {\em eg} the reviews~\cite{vacher2017a,kheifeits2023a}. 
 Our present study focuses on the orientation-resolved `molecular delays'  defined as
\begin{eqnarray}\label{eqn:rabdel}
\rabdel(\thetak)&=&\frac{\vartheta(\thetak)}{2\wIR}.
\end{eqnarray}
They express the angular variations of the so-called `atomic' (or `molecular') delays that were first introduced to interpret \rabbit experiments in terms of photoemission dynamics~\cite{klunder2011a}. 

\begin{figure}[t]
\includegraphics[height=\figheight]{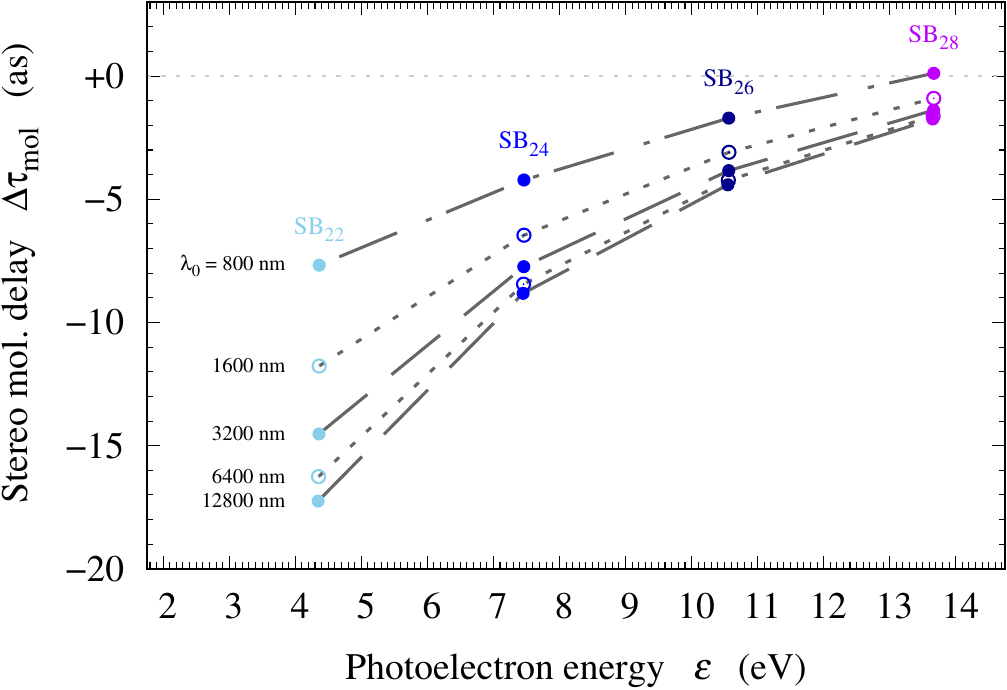}
\caption{\label{fig:rabdel_diff}  Stereo molecular delays $\Delta\rabdel$ (Eq~\ref{eqn:rabdel_diff}) as a function of the average sideband energy, computed for a set of \lIR values in \rabbit simulations, see caption of Fig.~\ref{fig:rabdel}.\signOK}
\end{figure}

We first present and analyze the results obtained with the standard fundamental wavelength, $\lIR=800$ nm. We considered odd harmonic orders ranging from $21$ to $29$, producing sideband photoelectrons near threshold ($\varepsilon<15$ eV). The orientation-resolved molecular delays on the left and right sides of the molecule are displayed  in frames (a) and (b) of Fig.~\ref{fig:rabdel} respectively (circles connected by dash-dotted guidelines, see wavelength labels). Their overall trend is typical of molecular delays in smooth continua, their magnitude decaying monotonically when the sideband energy increases. 
 Their values lie in the $100$ as range, with a relatively small anisotropy, which remarkably contrasts with the much shorter 1-photon counterparts displayed in Fig.~\ref{fig:wigdel}. The anisotropy is nevertheless significant at the attosecond time scale, consistently with former experimental and theoretical studies~\cite{chacon2014a,heuser2016a,hockett2017a,vos2018a}. This can be seen on Fig.~\ref{fig:rabdel_diff} where the stereo molecular delays
\begin{eqnarray}\label{eqn:rabdel_diff}
\Delta\rabdel&=&\rabdel(180\degree)-\rabdel(0\degree)
\end{eqnarray}
are plotted (same legend). Ranging within few tens of as at the lowest considered sideband, the magnitude of $\Delta\rabdel$ also follows a global decaying trend when $\varepsilon$ increases\footnote{That trend is not necessarily monotonic, \eg for $\lIR=800$ nm the delay  changes sign around \SB{28}.}. 
 They also significantly differ from the stereo Wigner delays computed at $\xref=\langle x \rangle_0$, in spite of lying within a similar as scale. Note that no alternative choice in the origin  provides a satisfactory agreement between the 1-photon and 2-photon stereo delays with these 800-nm \rabbit simulations, as will be further discussed in Section~\ref{sec:tauCC}.

\begin{table}
\begin{tabular}{|rl|rrrrr|}
\hline
\lIR & (nm) & $800$ & $1600$ & $3200$ & $6400$ & $12800$ \\
\hline\hline
$\vartheta(0\degree)$ & (rad) & -0.4872 & -0.3747 & -0.2661 & -0.1762 & -0.1097 \\
$\vartheta(180\degree)$ & (rad) & -0.5234 & -0.4024 & -0.2831 & -0.1858 & -0.1148 \\ \hline
$\rabdel(0\degree)$ & (as) & -103.4 & -159.0 & -225.8 & -299.0 & -372.6 \\
$\rabdel(180\degree)$ & (as) & -111.0 & -170.8 & -240.3 & -315.4 & -389.8 \\
\hline 
\end{tabular}
\caption{\label{tab:moldel} Orientation-resolved molecular phases $\vartheta(\thetak)$ and delays $\rabdel(\thetak)$ (Eq.~\ref{eqn:rabdel}) associated with $\SB{n\times22}$  for five values of the fundamental wavelength \lIR, $2^{n-1}\times800$ nm ($n=1-5$). These sidebands share the same average photoelectron energy, $\varepsilon=4.35$ eV.} 
\end{table}
As expected and in contrast to the Wigner delays studied in Sec.~\ref{sec:wigner}, the (orientation-resolved) molecular delays bear no intrinsic dependency with respect to the arbitrary origin \xref. Nevertheless, these measurable quantities depend on the `probe' laser wavelength $\lIR$~\cite{dahlstrom2013a,pazourek2015a}  
We illustrate this in Table~\ref{tab:moldel}, which displays the values of  $\rabdel(\thetak)$ obtained at \SB{n\times22}, on each side of the molecule, when the fundamental wavelength is varied following the geometric progression $2^{n-1}\times800$ nm ($n=1-5$)\footnote{Results for the largest considered wavelength (12800 nm) were obtained using 2\textsuperscript{nd} order perturbation theory rather than by solving the \tdse, to avoid numerical issues related to the tight spectral proximity of the photoelectron peaks.}. This \lIR sampling ensures all \SB{n\times22} sidebands to be centered at the same photoelectron energy than \SB{22} for the 800 nm case ($\varepsilon=4.35$ eV). It thus allows assessing the evolution of $\rabdel(\thetak)$ with respect to \lIR solely. Indeed, we observe here that the orientation-resolved molecular delays depend significantly on \lIR. That dependency is not only the result of the division by \wIR on the r.h.s of Eq.~\ref{eqn:rabdel}, since the orientation-resolved molecular phases $\vartheta(\thetak)$ themselves depend on \lIR in an antagonist way (see same Table and above mentioned references). 

The data obtained for higher sideband orders \SB{n\times(2q+1)} with the same set of wavelengths are also displayed in Figs.~\ref{fig:rabdel} and~\ref{fig:rabdel_diff}. The conclusions regarding the \lIR dependency are the same as for  \SB{n\times22}.
Focusing on that dependency, one should note that the orientation-resolved delay $\rabdel(\thetak)$ associated with each sideband energy, on each side of the molecule, follows an apparent logarithmic trend with respect to \lIR (In Fig.~\ref{fig:rabdel}, for each \SB{n\times 2q}, the values of $\rabdel(\thetak)$ are $\sim$ evenly spaced while \lIR follows a geometric progression). As will be discussed in Sec.~\ref{sec:tauCC}, this is consistent with previous studies and state of the art analytical derivations.  However, the \lIR-dependency {\em of their angular variations}, namely of the stereo molecular delays $\Delta\rabdel$ at fixed $\varepsilon$, is no longer logarithmic (see Fig.~\ref{fig:rabdel_diff}). Strikingly, it  instead appears to be converging when \lIR increases.
 
In the next section, we investigate numerically the link between the fundamental stereo Wigner delays and the measurable stereo molecular delays, by addressing the ways to account for the dependencies pertaining to each of them. 

\section{\rabbit as a pump-probe scheme to measure Wigner delays}\label{sec:tauCC}

When the \rabbit technique is used to investigate the dynamics of photoemission in terms of Wigner delays~\cite{klunder2011a}, it is seen as an interferometric pump-probe scheme where the \xuv harmonic pulse initiates  single-photon ionization processes, that are probed by the fundamental field at \lIR (typically in the \ir domain). Within this paradigm,  atomic (or molecular) delays are considered as measurements of  Wigner delays, modified by a correction term resulting from the \ir probe stage. Retrieving the former out of the latter thus requires the knowledge of the probe term, as introduced in~\cite{klunder2011a}. 
Here, we will see that the asymmetry of our model molecule induces angular variations of the molecular-Wigner delay differences that are significant on the attosecond scale. 

Using the orientation-dependent delays introduced in the previous sections, we define here the orientation-resolved probe delay as 
\begin{eqnarray}\label{eqn:cordel}
\cordel(\thetak)&=&\rabdel(\thetak)-\wigdel(\thetak).
\end{eqnarray}  
Below, we confront our orientation-averaged results to an available analytic model, before focusing on the details of their angular variations.

\subsection{Orientation averaged delays}

A convenient  closed-form expression modelling the \lIR-dependency of this correction was derived using asymptotic expansions and well delineated approximations, see~\cite{pazourek2015a}. 
With the notations of the present paper, it reads:
\begin{eqnarray}\label{eqn:correction_anal}
g_\varepsilon(\lIR)&=&\frac{Z}{(2\varepsilon)^{3/2}}\left[2-\ln(\varepsilon c\lIR)\right].
\end{eqnarray}
 $Z$ is in principle the charge associated with the asymptotic Coulomb field felt by the photoelectron, but it can also be used as an adjustable parameter compensating for some approximations. 
 This formula and related ones (see also~\cite{dahlstrom2013a}) are commonly  exploited as such in  experimental and theoretical studies, see \eg~\cite{huppert2016a,hockett2017a,gong2022a,boyer2023a}. 

The function given in Eq.~\ref{eqn:correction_anal} notably underlines the overall logarithmic \lIR-dependency of the molecular delays,  anticipated earlier when commenting our results displayed in Fig.~\ref{fig:rabdel}. Besides, it refers to Wigner delays defined with Coulomb reference waves, which is the natural choice for photoemission. However, it is an isotropic expression\footnote{Its isotropy comes from the hypothesis according to which the \lIR-field  probes the photoelectron outside the system- and channel-specific short range interaction zone, \ie where it feels only the universal -- isotropic -- Coulomb tail of the ionic potential.}. Hence, it cannot account for orientation-resolved measurements that reveal angular variations of the probe term, as \eg experimentally in He~\cite{heuser2016a} and theoretically in noble gaz atoms~\cite{bray2018a}, as well as in the present simulations, as developped below.

\begin{table}[t]
\begin{tabular}{rccccc}
\hline
$2q$ & : &$22$ & $24$ &  $26$ &  $28$ \\
$Z$ [err] & : & 0.80 [2\%] & 0.84 [2\%] & 0.85 [3\%] & 0.84 [5\%] \\
\hline
\end{tabular}
\caption{\label{tab:cordel_fits} Effective charge obtained by fitting the average probe delay $\avecordel$ (Eq.~\ref{eqn:correction_aver}) with the formula given in Eq.~\ref{eqn:correction_anal}, as a function of $\lIR$ at fixed $\varepsilon$ for each series of sidebands \SB{n\times2q} ($n$ parametrizes the \lIR sampling, see caption of Tab.~\ref{tab:moldel}). The standard relative error for each fit is indicated in brackets.
}
\end{table}
\begin{figure}[t]
\includegraphics[height=\figheight]{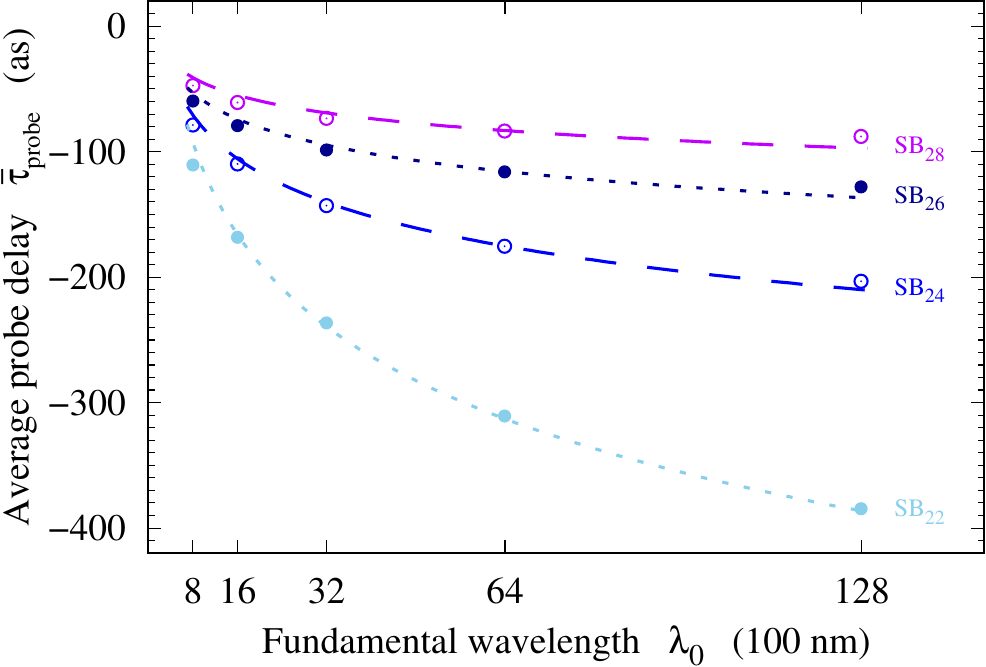}
\caption{
\label{fig:cordel_aver} Average probe delay against the \rabbit probe wavelength \lIR. Numerical results (Eq.~\ref{eqn:correction_aver}) are displayed as circles. Alternating dashed and dotted lines correspond to the universal closed-form expression (Eq.~\ref{eqn:correction_anal}), where $Z$ was used as a fitting parameter, see Table~\ref{tab:cordel_fits}. The results are displayed for the 4 sideband datasets considered all through the paper. For simplicity, each set is labelled with the sideband order in the 800 nm case. \signOK
}
\end{figure}

In order to assess the relevance of $g_\varepsilon(\lIR)$ in the context of the present work, we fitted it to the orientation averaged probe delays obtained in our numerical experiments
\begin{eqnarray}\label{eqn:correction_aver}
\avecordel&=&\frac{1}{2}\sum\limits_{\thetak=0\degree,180\degree} \cordel(\thetak),
\end{eqnarray} 
using the effective charge $Z$ as the sole fitting parameter. Note that the average value $\avecordel$ does {\em not} depend on the origin \xref, in contrast to the orientation-resolved counterpart $\cordel(\thetak)$\footnote{This is consistent with the results obtained in~\cite{huppert2016a}, where orientation-averaged simulations reached a satisfactory agreement with experiments without considering the origin issue.}. Indeed, the correction term inherited from $\wigdel(\thetak)$, modeled by Eq.~\ref{eqn:dwigdel}, cancels out when averaging over $\thetak$ (this can be verified \eg with the data of Tab.~\ref{tab:wigdel}). 

The fit was done independently for each sideband series \SB{n\times2q}, each of them corresponding to a fixed photoelectron energy $\varepsilon$. The four series provide consistent $Z$ values, lying around $0.83$ \signOK within the standard error of the fits ($\lesssim5\%$), see Tab.~\ref{tab:cordel_fits}. The average probe term \avecordel is displayed against \lIR in Fig.~\ref{fig:cordel_aver}, for the four considered series of sidebands. The fits, displayed on the same figure, appear to accurately follow the orientation-averaged numerical data, within the timescale of the plot.

We will now take a closer look at the angular variations of the probe delay which are not taken into account by the analytic model function $g_\varepsilon(\lIR)$, but nevertheless manifest at the attosecond time scale~\cite{liao2021a}.

\subsection{Reconciliating orientation-resolved molecular and Wigner delays}

\begin{figure*}
\includegraphics[width=2.1\figwidth]{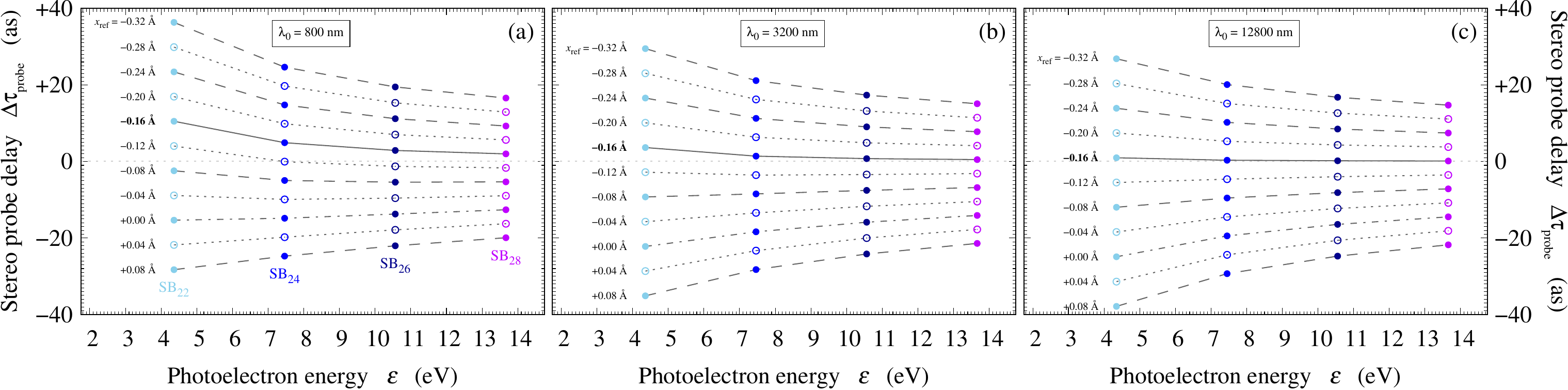}
\caption{\label{fig:cordel_diff} Stereo correction delay $\Delta\cordel$ (Eq.~\ref{eqn:cordel_diff}) as function of the photoelectron energy $\varepsilon$. Each frame (a-c) corresponds to a given value of \lIR used to measure the molecular delays. The data, shown as alternating filled and empty circles connected with guidelines, were computed at the energies of the sidebands considered in the \rabbit simulations.   Each line corresponds to a given value of \xref used to define and compute the Wigner delays. Solid lines correspond to the  datasets obtained with  $\xref=\langle x \rangle_0=-0.16$~\AA, which converge to 0 at all energies when \lIR increases.  Dash-dotted lines correspond to the datasets obtained with the arbitrary origin $\xref=0$. For simplicity, the sidebands are labelled for the 800~nm case only [frame (a)].} 
\end{figure*}

We eventually address the angular variation
\begin{eqnarray}\label{eqn:cordel_diff}
\Delta\cordel&=&\cordel(180\degree)-\cordel(0\degree) \\
                    &=&\Delta\rabdel-\Delta\wigdel,
\end{eqnarray}
hereafter referred to as the stereo probe delay. In contrast to the orientation-resolved or -averaged probe delays, it is not affected by the choice of the reference-wave used to define the Wigner delays. Instead, it inherits  the origin-dependency of the Wigner delay, see Sec.~\ref{sec:wigner}, and  the wavelength-dependency of the molecular delay, see Sec.~\ref{sec:rabbit} and above. 

The values of $\Delta\cordel$ obtained in our simulations are shown in Fig.~\ref{fig:cordel_diff} against the photoelectron energy~$\varepsilon$. Each frame (a-c) corresponds to a given probe wavelength \lIR value $2^{(n-1)}\times800$ nm ($n=1,3,5$). We performed a systematic scan of \xref and only show a selection of representative cases (each guideline corresponds to a given value of \xref).\footnote{ We have checked that the finite-difference approximation 
\begin{eqnarray}
\left.\frac{\partial \eta}{\partial \varepsilon}\right\vert_{\varepsilon_{2q}} 
\simeq 
\frac{\eta(\varepsilon_{2q}+\wIR)-\eta(\varepsilon_{2q}-\wIR)}{2\wIR}
\end{eqnarray}
which underlies any attempt to link  the atomic/molecular phases to the Wigner delays~\cite{klunder2011a}, remains accurate in all the considered cases. In the most pathological situation, \ie at the lowest considered energy probed with the largest laser frequency \wIR (\SB{22} at 800 nm), this approximation induces an error of few 0.1 as.}

Looking at frame (a), one can see that at 800 nm no \xref value allows the stereo probe delay to vanish at all energies -- the arbitrary coordinate origin $\xref=0$ and the initial average position $\xref=\langle x\rangle_0$ ($=-0.16$ \AA, see Eq.~\ref{eqn:x0}) making no exception. At this point, one may conclude that standard \rabbit measurements cannot provide arbitrarily accurate access to the angular variations of Wigner delays in molecular photoemission.

However, the results obtained at larger wavelengths single out the average initial position $\langle x \rangle_0$ as a specific $\xref$ origin for which $\Delta\cordel$ does strikingly vanish at all energies, when \lIR increases. This observation eventually comforts the idea that \rabbit  virtually probes molecular photoemission dynamics with high accuracy.  It also confirms the average initial electron position as a statistically representative parameter implicitly characterizing the dynamics studied by such experiments, as assumed in~\cite{vos2018a}. 

On the theory side, it provides an objective criterion for setting the electron position origin in anisotropic scattering delays computations. It is important here to keep in mind that  $\Delta\cordel$ as such (Eq.~\ref{eqn:cordel}) is not an experimentally measurable quantity. Its \xref-dependency directly issues from the {\em definition} of the  term $\Delta\wigdel$, and has nothing to do with the {\em measurement} of the term $\Delta\rabdel$. 
Significant values of $\Delta\cordel$ at large photoelectron energies are thus spurious signatures of an inappropriate reference position in the theoretical definitions of the (stereo) Wigner delay. They are therefore related to an ill defined clock  rather than to a physically relevant asymmetry in the photoemission dynamics. In contrast, non-zero $\Delta\wigdel$ persisting at low energies in standard 800 nm \rabbit measurements -- even though the position origin is properly set -- are signatures of the asymmetric probe influence on the measurements. They however vanish when the probe wavelength increases.

Hence, stereo Wigner delays are encoded in \rabbit measurements with attosecond resolution, when the position origin is properly set and in the limit of a vanishing probe frequency, or at sufficiently large photoelectron electron energies\footnote{We have reproduced the stereo Wigner delays displayed in Fig. 9(a) of reference~\cite{liao2021a} and verified that a proper correction of the origin brings the data significantly closer to the 800 nm-\rabbit delays reported in the same figure, in particular at large energies.}. Their angular variations are related to differences in the potential energy landscape explored by the photoelectron while escaping the molecule in one direction or the other, and cannot be directly related to the `technical', potential-independant, correction given in Eq.~\ref{eqn:dwigdel}. This partly contradicts the interpretations of~\cite{vos2018a,liao2021a}, without questioning the main conclusions of these studies.

On the experimental side, the probe wavelengths cannot be arbitrarily increased in practice as it is done in the present simulations. The question of the magnitude of $\Delta\cordel$ thus remains to be dealt with, even when properly setting the position origin\footnote{All through the remain of the paper, the choice $\xref=\langle x \rangle_0$ is assumed.}. That magnitude is expected to highly depend on the degree of asymmetry of the probed molecule, among other things. 
Deriving a universal $\lIR-$ and $\varepsilon-$dependent model formula for $\Delta\cordel$ is beyond the scope of the present paper. In this last part, we will review what can be learned from our simulations with respect to this probe term.

First, one should note that in the present case $\Delta\cordel$ is relatively small, even for $\lIR=800$ nm. In the considered near-threshold energy range, it harldly exceeds 10~as around 4~eV and then decays monotonically, reaching $\sim 2$~as before 14~eV. 
We can thus expect the stereo probe delay to quickly reach the sub-as time scale, which should make it negligible when considering photoemission at  much higher energies, and when the molecular asymmetry is not too pronounced. This includes the results published in~\cite{chacon2014a} where stereo Wigner delays were investigated numerically in the context of streaking   rather than \rabbit. As mentionned earlier in the introduction, these results show no significant difference between the Wigner and streaking delays beyond $\sim20$ eV.

\begin{figure}[t]
\includegraphics[width=0.9\figwidth]{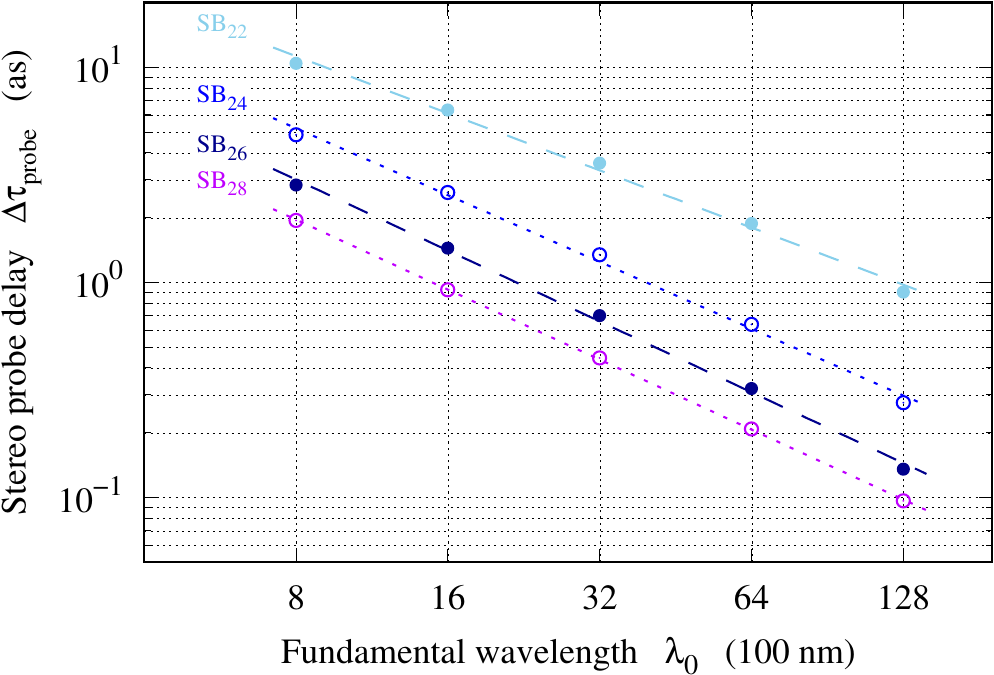}
\caption{\label{fig:cordel_fits} Stereo probe delay (Eq.~\ref{eqn:cordel_diff}) as a function of the probe wavelength \lIR, for each set of sidebands centered at a given energy (each set is labelled with the sideband order of the 800 nm case). The numerical data obtained in our simulations are indicated as symbols (same legend as in Fig.~\ref{fig:cordel_aver}). The fits of the function $h(\lIR)=a\times\lIR^{-p}$ to each set of data are shown as lines (alternating dashed and dotted styles). The fitted values of the exponent $p$ are provided in Tab.~\ref{tab:cordif_fits}. The plot is displayed in a log-log scale.}
\end{figure}We also investigated the evolution of $\Delta\cordel$ against the probe wavelength, for each of the considered photoelectron energies. The results are displayed in Fig.~\ref{fig:cordel_fits}. 
The striking alignements of the data, for each energy, in this log-log plot suggest a general $\lIR$-dependency following simple decaying power laws, as confirmed  by fitting the data with a $\lIR^{-p}$ law. We obtained $p$ values around $1$ slowly increasing with the photoelectron energy, see Table~\ref{tab:cordif_fits}. 
We performed additional \threed simulations on the H atom (not shown here), where we measured orientation-resolved atomic delays with \rabbit. We found that their angular variations, when they are significant, also display \lIR-dependencies following similar power laws, which stabilize, when \lIR increases, towards exponent values within $\sim 0.75 - 1.5$ depending on the orientation. In that case, the angular (momentum) dependency of the probe delay term is attributed to the $\ell$-dependent $\sim 1/r^2$ centrifugal barrier~\cite{fuchs2020a}, the plain atomic potential being otherwise spherically symmetric. These empirical, numerical findings suggest the existence of an underlying analytical law for the angular variations of $\Delta\cordel$ depending on the molecule asymmetry, the photoelectron energy and the probe laser wavelength. Such an analytical law remains to be established.
\begin{table}[t]
\begin{tabular}{rccccc}
\hline
$2q$ & : &$22$ & $24$ &  $26$ &  $28$ \\
$p$ [err] & : & 0.88 [5\%] & 1.03 [4\%] & 1.09 [3\%] & 1.08 [1\%] \\
\hline
\end{tabular}
\caption{\label{tab:cordif_fits} Power laws obtained by fitting the stereo probe delay $\Delta\cordel$ (Eq.~\ref{eqn:cordel_diff}) with $h(\lIR)=a\times\lIR^{-p}$, at fixed $\varepsilon$ for each series of sidebands \SB{n\times2q} ($n$ parametrizes the \lIR sampling, see caption of Table~\ref{tab:moldel}). The standard relative error for each fit is indicated in brackets.
}
\end{table}

\section{Summary and conclusions}
Using numerical experiments, we reviewed the physics underlying the concepts of Wigner delay, that characterize single photon ionization processes, and of so-called atomic or molecular delay inferred from \rabbit measurements, their interpretations and their connections.

The simulations were performed on a simple \oned asymmetric model molecule, with a particular focus on the angular variations of the delays with respect to the photoelectron ejection direction, and on the intrinsic dependencies of the delays with respect to theoretical and experimental parameters.
 
We verified that our orientation-averaged results are consistent with an available analytical formula~\cite{pazourek2015a} for the difference between the molecular and Wigner delays. By construction, that formula accounts for the wavelength dependency of the molecular delay on the one hand, and for the specific choice of Coulomb reference waves to define and compute the Wigner delays on the other hand. Yet, it is an isotropic formula that cannot account for the angular variations of the molecular and Wigner delays, nor of their differences. 

Addressing this issue puts forward the intrinsic dependency of the orientation-resolved Wigner delays with respect to an arbitrary electron position origin -- which does not concern molecular delays. Our simulations however show that orientation-resolved molecular phases and Wigner delays can be related, when and only when the latter is defined with a {\it specific origin corresponding to the average electron position in the initial bound state}. This demonstrates empirically that the initial average electron position is encoded in \rabbit measurements, and thus provides an unambiguous prescription for setting the electron position origin in the definition and computation of orientation-resolved Wigner delays. {A different position origin, when not corrected, can lead to dramatically and spuriously large delays even when it is set `within' the molecule, \ie few tenths of \AA\ from the average initial position. The magnitude of such artificially enhanced delays is then likely to hide the actual, physically relevant, anisotropic dynamics occurring at the attosecond scale.

Moreover, our simulations reveal angular variations of the molecular delays matching those of the Wigner delays {\it in the limit of a vanishing \rabbit probe frequency}.  The wavelengths considered in our simulations extend way beyond those routinely available in real-life experiments. However, the differences observed even at 800 nm are rather small. They also decay rapidly when the photoelectron energy increases. Therefore, one can expect them to be negligible in many practical cases, depending on the photoelectron energy and the degree of asymmetry of the molecule. Finally, the observed trend with respect to \lIR will serve as a basis for further analytical developments in view to include a proper angular dependency in the probe term, including using more elaborate field configurations~\cite{wang2022a,boll2023a,kheifets2023b,han2023a}. This will allow observing the dynamics of a most fundamental quantum process in all its complexity, with actual attosecond accuracy.

\begin{acknowledgments}
The authors acknowledge fruitful and stimulating discussions with Pascal Sali\`eres and Bernard Pons. This research received the financial support of the French National Research Agency through Grants No. ANR-15-CE30-0001-CIMBAAD and ANR-20-CE30-0007-DECAP.
\end{acknowledgments}


\appendix
\section{Partial-wave expansion in \oned}\label{sec:pwexpand}

\subsection{\oned polar coordinates}
In \oned, position is commonly characterized by a single coordinate $z\in\mathbb{R}$. Alternatively, it can be expressed {\em via} a set of two `polar' coordinates consisting in
\begin{eqnarray}
\begin{cases}
 \text{\ a radial coordinate} & r=\vert z \vert\ (r\in\mathbb{R}^+)\ ; \\
 \text{\  an angular coordinate} & \theta=\arccos(z/\vert z\vert)\ (\theta\in\{0,\pi\}).
\end{cases}
\end{eqnarray}
In spite of being unusual, this representation is of interest to highlight existing analogies between \oned scattering problems~\cite{saalmann2023a,eberly1965a,*formanek1976a,*sassoli1994a,*nogami1996a,*barlette2001a,*boya2008a,*elghazawy2023a} and problems in higher dimensions where partial-wave expansions are standard.

\subsection{Angular basis functions}
In analogy with \threed spherical harmonics, let us introduce the following set of basis functions: 
\begin{eqnarray}
Y_{\ell}(\theta)&=&\frac{1}{\sqrt{2}}{(\cos\theta)^\ell}, \ell\in\{0,1\}
\end{eqnarray}
for the discrete angular coordinate $\theta$ introduced in Eq.~\ref{eqn:polar_def}.
Defined as such, the functions are orthogonal and normalized to unity,
\begin{eqnarray}\label{eqn:orth}
( Y_\ell \vert Y_{\ell'} ) &=& \kro{\ell}{\ell'}
\end{eqnarray}
where \kro{}{} is the Kronecker symbol and the $(\ \vert\ )$ bra-ket notation refers to the  scalar product with respect to the angular variable:
\begin{eqnarray}
( A \vert B ) &=& \sum\limits_{\theta=0,\pi} [A(\theta)]^\star 
B(\theta).
\end{eqnarray}  
The two functions further verify the following composition rule:
\begin{eqnarray}\label{eqn:comp}
\nonumber Y_\ell(\theta) Y_{\ell'}(\theta)&=&\sqrt{2}\left[\kro{\ell}{\ell'}\times Y_{0}(\theta) +(1-\kro{\ell}{\ell'})\times Y_{1}(\theta)\right]. \\
\end{eqnarray}
One must note that the values assigned to the index $\ell$ are arbitrary. In particular, this `quantum number' cannot be associated with an angular momentum since  the angular coordinate is not continuous (see Noether's theorem~\cite{goldstein1980a}) -- \ie rotation is not defined in \oned. Yet, it is related to the parity properties of \oned functions, as exploited in the following.

\subsection{Partial waves}
The even ($f_{0}$) and odd ($f_{1}$) components of any function $F$ can be defined as
\begin{eqnarray}
f_{\ell}(r)&=&\frac{1}{\sqrt{2}}\left[F(r)+(-1)^\ell F(-r)\right] 
\end{eqnarray}
for all $r \in \mathbb{R}^+$.
The normalisation factor $1/\sqrt{2}$ is here introduced such that $F(z)$ expressed with the  $(r,\theta)$ coordinates takes the simple form of a partial-wave expansion:
\begin{eqnarray}
F(z)&=&\sum\limits_{\ell=0,1} f_\ell(r) Y_\ell(\theta).
\end{eqnarray}
Hence, $Y_{0}(\theta)$ corresponds to the angular part of an even function of $z$, and $Y_{1}(\theta)$ to the one of an odd function.

\subsection{Schr\"odinger equation}
We consider the time-independent Schr\"odinger equation 
\begin{eqnarray}\label{eqn:schropartial}
\left[-\frac{1}{2}\frac{\der^2}{\der z^2}+V(z)-\ener\right]\Psi(z)&=&0.
\end{eqnarray}
of a \oned particle (of mass equal to 1)  with total energy \ener, where the \oned partial-wave expansions of the potential and of the solution wave-function read:
\begin{eqnarray}
\label{eqn:radVne} V(z)&=&\sum\limits_{\ell=0,1} v_\ell(r) Y_\ell(\theta), \\
\label{eqn:radPsi} \Psi(z)&=&\sum\limits_{\ell=0,1}\psi_{\ell}(r) Y_\ell(\theta).
\end{eqnarray}
Projecting the Schr\"odinger equation~\ref{eqn:schropartial} on the angular basis  $\{Y_0(\theta),Y_1(\theta)\}$, and using Eqs.~\ref{eqn:orth},~\ref{eqn:comp} and 
\begin{eqnarray}
\left( Y_\ell \left\vert \frac{\der^2}{\der z^2} \right\vert Y_{\ell'} \right)&=&\kro{\ell}{\ell'}\frac{\der^2}{\der r^2},
\end{eqnarray}
one gets the following coupled equations for the radial functions $\psi_{\ell}(r)$~:
\begin{eqnarray}
\nonumber \left[-\frac{1}{2}\frac{\der^2}{\der r^2}+v_{0}(r)-\varepsilon\right]\psi_{\ell}(r)+v_{1}(r)\psi_{1-\ell}(r)&=&0 \\
\label{eqn:coupledschro}
\end{eqnarray}
where the odd component of the potential $v_{-1}(r)$ is responsible for the coupling between the two partial waves $\psi_{0}(r)$ and $\psi_{1}(r)$. 

The continuity conditions of the solution wave-function $\Psi(z)$ translate into the following boundary conditions at the origin:
\begin{eqnarray}
\psi'_{0}(0)&=&0, \\
\psi_{1}(0)&=&0,
\end{eqnarray}
to be imposed when implementing the numerical resolution of the system (Eq.~\ref{eqn:coupledschro}). In the doubly degenerate continuum,  two linearly independent wave-functions can be obtained by solving the system with two linearly independant arbitrary sets of initial conditions $\{\psi_{0}(0),\psi'_{1}(0)\}$. 


\begin{thebibliography}{58}%
\makeatletter
\providecommand \@ifxundefined [1]{%
 \@ifx{#1\undefined}
}%
\providecommand \@ifnum [1]{%
 \ifnum #1\expandafter \@firstoftwo
 \else \expandafter \@secondoftwo
 \fi
}%
\providecommand \@ifx [1]{%
 \ifx #1\expandafter \@firstoftwo
 \else \expandafter \@secondoftwo
 \fi
}%
\providecommand \natexlab [1]{#1}%
\providecommand \enquote  [1]{``#1''}%
\providecommand \bibnamefont  [1]{#1}%
\providecommand \bibfnamefont [1]{#1}%
\providecommand \citenamefont [1]{#1}%
\providecommand \href [0]{\begingroup \@sanitize@url \@href}%
\providecommand \@href[1]{\@@startlink{#1}\@@href}%
\providecommand \@@href[1]{\endgroup#1\@@endlink}%
\providecommand \@sanitize@url [0]{\catcode `\\12\catcode `\$12\catcode
  `\&12\catcode `\#12\catcode `\^12\catcode `\_12\catcode `\%12\relax}%
\providecommand \@@startlink[1]{}%
\providecommand \@@endlink[0]{}%
\providecommand \url  [0]{\begingroup\@sanitize@url \@url }%
\providecommand \@url [1]{\endgroup\@href {#1}{\urlprefix }}%
\providecommand \urlprefix  [0]{URL }%
\providecommand \Eprint [0]{\href }%
\providecommand \doibase [0]{http://dx.doi.org/}%
\providecommand \selectlanguage [0]{\@gobble}%
\providecommand \bibinfo  [0]{\@secondoftwo}%
\providecommand \bibfield  [0]{\@secondoftwo}%
\providecommand \translation [1]{[#1]}%
\providecommand \BibitemOpen [0]{}%
\providecommand \bibitemStop [0]{}%
\providecommand \bibitemNoStop [0]{.\EOS\space}%
\providecommand \EOS [0]{\spacefactor3000\relax}%
\providecommand \BibitemShut  [1]{\csname bibitem#1\endcsname}%
\let\auto@bib@innerbib\@empty
\bibitem [{\citenamefont {Cavalieri}\ \emph {et~al.}(2007)\citenamefont
  {Cavalieri}, \citenamefont {Muller}, \citenamefont {Uphues}, \citenamefont
  {Yakovlev}, \citenamefont {Baltu{\v s}ka}, \citenamefont {Horvath},
  \citenamefont {Schmidt}, \citenamefont {Blumel}, \citenamefont {Holzwarth},
  \citenamefont {Hendel}, \citenamefont {Drescher}, \citenamefont {Kleineberg},
  \citenamefont {Echenique}, \citenamefont {Kienberger}, \citenamefont
  {Krausz},\ and\ \citenamefont {Heinzmann}}]{cavalieri2007a}%
  \BibitemOpen
  \bibfield  {author} {\bibinfo {author} {\bibfnamefont {A.~L.}\ \bibnamefont
  {Cavalieri}}, \bibinfo {author} {\bibfnamefont {N.}~\bibnamefont {Muller}},
  \bibinfo {author} {\bibfnamefont {T.}~\bibnamefont {Uphues}}, \bibinfo
  {author} {\bibfnamefont {V.~S.}\ \bibnamefont {Yakovlev}}, \bibinfo {author}
  {\bibfnamefont {A.}~\bibnamefont {Baltu{\v s}ka}}, \bibinfo {author}
  {\bibfnamefont {B.}~\bibnamefont {Horvath}}, \bibinfo {author} {\bibfnamefont
  {B.}~\bibnamefont {Schmidt}}, \bibinfo {author} {\bibfnamefont
  {L.}~\bibnamefont {Blumel}}, \bibinfo {author} {\bibfnamefont
  {R.}~\bibnamefont {Holzwarth}}, \bibinfo {author} {\bibfnamefont
  {S.}~\bibnamefont {Hendel}}, \bibinfo {author} {\bibfnamefont
  {M.}~\bibnamefont {Drescher}}, \bibinfo {author} {\bibfnamefont
  {U.}~\bibnamefont {Kleineberg}}, \bibinfo {author} {\bibfnamefont {P.~M.}\
  \bibnamefont {Echenique}}, \bibinfo {author} {\bibfnamefont {R.}~\bibnamefont
  {Kienberger}}, \bibinfo {author} {\bibfnamefont {F.}~\bibnamefont {Krausz}},
  \ and\ \bibinfo {author} {\bibfnamefont {U.}~\bibnamefont {Heinzmann}},\
  }\bibfield  {title} {\enquote {\bibinfo {title} {{Attosecond spectroscopy in
  condensed matter}},}\ } {\bibfield  {journal} {\bibinfo
  {journal} {Nature}\ }\textbf {\bibinfo {volume} {449}},\ \bibinfo {pages}
  {1029} (\bibinfo {year} {2007})}\BibitemShut {NoStop}%
\bibitem [{\citenamefont {Schultze}\ \emph {et~al.}(2010)\citenamefont
  {Schultze}, \citenamefont {Fie{\ss}}, \citenamefont {Karpowicz},
  \citenamefont {Gagnon}, \citenamefont {Korbman}, \citenamefont {Hofstetter},
  \citenamefont {Neppl}, \citenamefont {Cavalieri}, \citenamefont {Komninos},
  \citenamefont {Mercouris}, \citenamefont {Nicolaides}, \citenamefont
  {Pazourek}, \citenamefont {Nagele}, \citenamefont {Feist}, \citenamefont
  {Burgd{\"o}rfer}, \citenamefont {Azzeer}, \citenamefont {Ernstorfer},
  \citenamefont {Kienberger}, \citenamefont {Kleineberg}, \citenamefont
  {Goulielmakis}, \citenamefont {Krausz},\ and\ \citenamefont
  {Yakovlev}}]{schultze2010a}%
  \BibitemOpen
  \bibfield  {author} {\bibinfo {author} {\bibfnamefont {M.}~\bibnamefont
  {Schultze}}, \bibinfo {author} {\bibfnamefont {M.}~\bibnamefont {Fie{\ss}}},
  \bibinfo {author} {\bibfnamefont {N.}~\bibnamefont {Karpowicz}}, \bibinfo
  {author} {\bibfnamefont {J.}~\bibnamefont {Gagnon}}, \bibinfo {author}
  {\bibfnamefont {M.}~\bibnamefont {Korbman}}, \bibinfo {author} {\bibfnamefont
  {M.}~\bibnamefont {Hofstetter}}, \bibinfo {author} {\bibfnamefont
  {S.}~\bibnamefont {Neppl}}, \bibinfo {author} {\bibfnamefont {A.~L.}\
  \bibnamefont {Cavalieri}}, \bibinfo {author} {\bibfnamefont {Y.}~\bibnamefont
  {Komninos}}, \bibinfo {author} {\bibfnamefont {T.}~\bibnamefont {Mercouris}},
  \bibinfo {author} {\bibfnamefont {C.~A.}\ \bibnamefont {Nicolaides}},
  \bibinfo {author} {\bibfnamefont {R.}~\bibnamefont {Pazourek}}, \bibinfo
  {author} {\bibfnamefont {S.}~\bibnamefont {Nagele}}, \bibinfo {author}
  {\bibfnamefont {J.}~\bibnamefont {Feist}}, \bibinfo {author} {\bibfnamefont
  {J.}~\bibnamefont {Burgd{\"o}rfer}}, \bibinfo {author} {\bibfnamefont
  {A.~M.}\ \bibnamefont {Azzeer}}, \bibinfo {author} {\bibfnamefont
  {R.}~\bibnamefont {Ernstorfer}}, \bibinfo {author} {\bibfnamefont
  {R.}~\bibnamefont {Kienberger}}, \bibinfo {author} {\bibfnamefont
  {U.}~\bibnamefont {Kleineberg}}, \bibinfo {author} {\bibfnamefont
  {E.}~\bibnamefont {Goulielmakis}}, \bibinfo {author} {\bibfnamefont
  {F.}~\bibnamefont {Krausz}}, \ and\ \bibinfo {author} {\bibfnamefont {V.~S.}\
  \bibnamefont {Yakovlev}},\ }\bibfield  {title} {\enquote {\bibinfo {title}
  {Delay in photoemission},}\ } {} {\bibfield  {journal} {\bibinfo
  {journal} {Science}\ }\textbf {\bibinfo {volume} {328}},\ \bibinfo {pages}
  {1658} (\bibinfo {year} {2010})}\BibitemShut {NoStop}%
\bibitem [{\citenamefont {Haessler}\ \emph {et~al.}(2009)\citenamefont
  {Haessler}, \citenamefont {Fabre}, \citenamefont {Higuet}, \citenamefont
  {Caillat}, \citenamefont {Ruchon}, \citenamefont {Breger}, \citenamefont
  {Carr\'e}, \citenamefont {Constant}, \citenamefont {Maquet}, \citenamefont
  {M\'evel}, \citenamefont {Sali\`eres}, \citenamefont {Ta{\"\i}eb},\ and\
  \citenamefont {Mairesse}}]{haessler2009a}%
  \BibitemOpen
  \bibfield  {author} {\bibinfo {author} {\bibfnamefont {S.}~\bibnamefont
  {Haessler}}, \bibinfo {author} {\bibfnamefont {B.}~\bibnamefont {Fabre}},
  \bibinfo {author} {\bibfnamefont {J.}~\bibnamefont {Higuet}}, \bibinfo
  {author} {\bibfnamefont {J.}~\bibnamefont {Caillat}}, \bibinfo {author}
  {\bibfnamefont {T.}~\bibnamefont {Ruchon}}, \bibinfo {author} {\bibfnamefont
  {P.}~\bibnamefont {Breger}}, \bibinfo {author} {\bibfnamefont
  {B.}~\bibnamefont {Carr\'e}}, \bibinfo {author} {\bibfnamefont
  {E.}~\bibnamefont {Constant}}, \bibinfo {author} {\bibfnamefont
  {A.}~\bibnamefont {Maquet}}, \bibinfo {author} {\bibfnamefont
  {E.}~\bibnamefont {M\'evel}}, \bibinfo {author} {\bibfnamefont
  {P.}~\bibnamefont {Sali\`eres}}, \bibinfo {author} {\bibfnamefont
  {R.}~\bibnamefont {Ta{\"\i}eb}}, \ and\ \bibinfo {author} {\bibfnamefont
  {Y.}~\bibnamefont {Mairesse}},\ }\bibfield  {title} {\enquote {\bibinfo
  {title} {Phase-resolved attosecond near-threshold photoionization of
  molecular nitrogen},}\ }\href {\doibase 10.1103/PhysRevA.80.011404}
  {\bibfield  {journal} {\bibinfo  {journal} {Phys. Rev. A}\ }\textbf {\bibinfo
  {volume} {80}},\ \bibinfo {pages} {011404} (\bibinfo {year}
  {2009})}\BibitemShut {NoStop}%
\bibitem [{\citenamefont {Kl{\"u}nder}\ \emph {et~al.}(2011)\citenamefont
  {Kl{\"u}nder}, \citenamefont {Dahlstr{\"o}m}, \citenamefont {Gisselbrecht},
  \citenamefont {Fordell}, \citenamefont {Swoboda}, \citenamefont {Gu\'enot},
  \citenamefont {Johnsson}, \citenamefont {Caillat}, \citenamefont
  {Mauritsson}, \citenamefont {Maquet}, \citenamefont {Ta{\"\i}eb},\ and\
  \citenamefont {L'Huillier}}]{klunder2011a}%
  \BibitemOpen
  \bibfield  {author} {\bibinfo {author} {\bibfnamefont {K.}~\bibnamefont
  {Kl{\"u}nder}}, \bibinfo {author} {\bibfnamefont {J.~M.}\ \bibnamefont
  {Dahlstr{\"o}m}}, \bibinfo {author} {\bibfnamefont {M.}~\bibnamefont
  {Gisselbrecht}}, \bibinfo {author} {\bibfnamefont {T.}~\bibnamefont
  {Fordell}}, \bibinfo {author} {\bibfnamefont {M.}~\bibnamefont {Swoboda}},
  \bibinfo {author} {\bibfnamefont {D.}~\bibnamefont {Gu\'enot}}, \bibinfo
  {author} {\bibfnamefont {P.}~\bibnamefont {Johnsson}}, \bibinfo {author}
  {\bibfnamefont {J.}~\bibnamefont {Caillat}}, \bibinfo {author} {\bibfnamefont
  {J.}~\bibnamefont {Mauritsson}}, \bibinfo {author} {\bibfnamefont
  {A.}~\bibnamefont {Maquet}}, \bibinfo {author} {\bibfnamefont
  {R.}~\bibnamefont {Ta{\"\i}eb}}, \ and\ \bibinfo {author} {\bibfnamefont
  {A.}~\bibnamefont {L'Huillier}},\ }\bibfield  {title} {\enquote {\bibinfo
  {title} {Probing single-photon ionization on the attosecond time scale},}\
  } {} {\bibfield  {journal} {\bibinfo  {journal} {Phys. Rev. Lett.}\
  }\textbf {\bibinfo {volume} {106}},\ \bibinfo {pages} {143002} (\bibinfo
  {year} {2011})}\BibitemShut {NoStop}%
\bibitem [{\citenamefont {Wigner}(1955)}]{wigner1955a}%
  \BibitemOpen
  \bibfield  {author} {\bibinfo {author} {\bibfnamefont {E.~P.}\ \bibnamefont
  {Wigner}},\ }\bibfield  {title} {\enquote {\bibinfo {title} {Lower limit for
  the energy derivative of the scattering phase shift},}\ }\href {\doibase
  10.1103/PhysRev.98.145} {\bibfield  {journal} {\bibinfo  {journal} {Phys.
  Rev.}\ }\textbf {\bibinfo {volume} {98}},\ \bibinfo {pages} {145} (\bibinfo
  {year} {1955})}\BibitemShut {NoStop}%
\bibitem [{\citenamefont {Muller}(2002)}]{muller2002a}%
  \BibitemOpen
  \bibfield  {author} {\bibinfo {author} {\bibfnamefont {H.~G.}\ \bibnamefont
  {Muller}},\ }\bibfield  {title} {\enquote {\bibinfo {title} {Reconstruction
  of attosecond harmonic beating by interference of two-photon transitions},}\
  }\href {\doibase 10.1007/s00340-002-0894-8} {\bibfield  {journal} {\bibinfo
  {journal} {Applied Physics B}\ }\textbf {\bibinfo {volume} {74}} (\bibinfo
  {year} {2002}),\ 10.1007/s00340-002-0894-8}\BibitemShut {NoStop}%
\bibitem [{\citenamefont {V\'eniard}\ \emph {et~al.}(1996)\citenamefont
  {V\'eniard}, \citenamefont {Ta\"ieb},\ and\ \citenamefont
  {Maquet}}]{veniard1996a}%
  \BibitemOpen
  \bibfield  {author} {\bibinfo {author} {\bibfnamefont {V.}~\bibnamefont
  {V\'eniard}}, \bibinfo {author} {\bibfnamefont {R.}~\bibnamefont {Ta\"ieb}},
  \ and\ \bibinfo {author} {\bibfnamefont {A.}~\bibnamefont {Maquet}},\
  }\bibfield  {title} {\enquote {\bibinfo {title} {Phase dependence of
  (\textit{N}+1)-color (\textit{N}$>$1) ir-uv photoionization of atoms with
  higher harmonics},}\ }\href {http://link.aps.org/doi/10.1103/PhysRevA.54.721}
  {\bibfield  {journal} {\bibinfo  {journal} {Phys. Rev. A}\ }\textbf {\bibinfo
  {volume} {54}},\ \bibinfo {pages} {721} (\bibinfo {year} {1996})}\BibitemShut
  {NoStop}%
\bibitem [{\citenamefont {Paul}\ \emph {et~al.}(2001)\citenamefont {Paul},
  \citenamefont {Toma}, \citenamefont {Breger}, \citenamefont {Mullot},
  \citenamefont {Aug\'e}, \citenamefont {Balcou}, \citenamefont {Muller},\ and\
  \citenamefont {Agostini}}]{paul2001a}%
  \BibitemOpen
  \bibfield  {author} {\bibinfo {author} {\bibfnamefont {P.~M.}\ \bibnamefont
  {Paul}}, \bibinfo {author} {\bibfnamefont {E.~S.}\ \bibnamefont {Toma}},
  \bibinfo {author} {\bibfnamefont {P.}~\bibnamefont {Breger}}, \bibinfo
  {author} {\bibfnamefont {G.}~\bibnamefont {Mullot}}, \bibinfo {author}
  {\bibfnamefont {F.}~\bibnamefont {Aug\'e}}, \bibinfo {author} {\bibfnamefont
  {P.}~\bibnamefont {Balcou}}, \bibinfo {author} {\bibfnamefont {H.~G.}\
  \bibnamefont {Muller}}, \ and\ \bibinfo {author} {\bibfnamefont
  {P.}~\bibnamefont {Agostini}},\ }\bibfield  {title} {\enquote {\bibinfo
  {title} {Observation of a train of attosecond pulses from high harmonic
  generation},}\ } {} {\bibfield  {journal} {\bibinfo  {journal}
  {Science}\ }\textbf {\bibinfo {volume} {292}},\ \bibinfo {pages} {1689}
  (\bibinfo {year} {2001})}\BibitemShut {NoStop}%
\bibitem [{\citenamefont {Mairesse}\ \emph {et~al.}(2003)\citenamefont
  {Mairesse}, \citenamefont {de~Bohan}, \citenamefont {Frasinski},
  \citenamefont {Merdji}, \citenamefont {Dinu}, \citenamefont {Monchicourt},
  \citenamefont {Breger}, \citenamefont {Kova{\v c}ev}, \citenamefont
  {Ta{\"\i}eb}, \citenamefont {Carr\'e}, \citenamefont {Muller}, \citenamefont
  {Agostini},\ and\ \citenamefont {Sali\`eres}}]{mairesse2003a}%
  \BibitemOpen
  \bibfield  {author} {\bibinfo {author} {\bibfnamefont {Y.}~\bibnamefont
  {Mairesse}}, \bibinfo {author} {\bibfnamefont {A.}~\bibnamefont {de~Bohan}},
  \bibinfo {author} {\bibfnamefont {L.~J.}\ \bibnamefont {Frasinski}}, \bibinfo
  {author} {\bibfnamefont {H.}~\bibnamefont {Merdji}}, \bibinfo {author}
  {\bibfnamefont {L.~C.}\ \bibnamefont {Dinu}}, \bibinfo {author}
  {\bibfnamefont {P.}~\bibnamefont {Monchicourt}}, \bibinfo {author}
  {\bibfnamefont {P.}~\bibnamefont {Breger}}, \bibinfo {author} {\bibfnamefont
  {M.}~\bibnamefont {Kova{\v c}ev}}, \bibinfo {author} {\bibfnamefont
  {R.}~\bibnamefont {Ta{\"\i}eb}}, \bibinfo {author} {\bibfnamefont
  {B.}~\bibnamefont {Carr\'e}}, \bibinfo {author} {\bibfnamefont {H.~G.}\
  \bibnamefont {Muller}}, \bibinfo {author} {\bibfnamefont {P.}~\bibnamefont
  {Agostini}}, \ and\ \bibinfo {author} {\bibfnamefont {P.}~\bibnamefont
  {Sali\`eres}},\ }\bibfield  {title} {\enquote {\bibinfo {title} {Attosecond
  synchronization of high-harmonic soft x-rays},}\ } {} {\bibfield
  {journal} {\bibinfo  {journal} {Science}\ }\textbf {\bibinfo {volume}
  {302}},\ \bibinfo {pages} {1540} (\bibinfo {year} {2003})}\BibitemShut
  {NoStop}%
\bibitem [{\citenamefont {Hentschel}\ \emph {et~al.}(2001)\citenamefont
  {Hentschel}, \citenamefont {Kienberger}, \citenamefont {Spielmann},
  \citenamefont {Reider}, \citenamefont {Milosevic}, \citenamefont {Brabec},
  \citenamefont {Corkum}, \citenamefont {Heinzmann}, \citenamefont {Drescher},\
  and\ \citenamefont {Krausz}}]{hentschel2001a}%
  \BibitemOpen
  \bibfield  {author} {\bibinfo {author} {\bibfnamefont {M.}~\bibnamefont
  {Hentschel}}, \bibinfo {author} {\bibfnamefont {R.}~\bibnamefont
  {Kienberger}}, \bibinfo {author} {\bibfnamefont {C.}~\bibnamefont
  {Spielmann}}, \bibinfo {author} {\bibfnamefont {G.~A.}\ \bibnamefont
  {Reider}}, \bibinfo {author} {\bibfnamefont {N.}~\bibnamefont {Milosevic}},
  \bibinfo {author} {\bibfnamefont {T.}~\bibnamefont {Brabec}}, \bibinfo
  {author} {\bibfnamefont {P.}~\bibnamefont {Corkum}}, \bibinfo {author}
  {\bibfnamefont {U.}~\bibnamefont {Heinzmann}}, \bibinfo {author}
  {\bibfnamefont {M.}~\bibnamefont {Drescher}}, \ and\ \bibinfo {author}
  {\bibfnamefont {F.}~\bibnamefont {Krausz}},\ }\bibfield  {title} {\enquote
  {\bibinfo {title} {Attosecond metrology},}\ } {} {\bibfield
  {journal} {\bibinfo  {journal} {Nature}\ }\textbf {\bibinfo {volume} {414}},\
  \bibinfo {pages} {509} (\bibinfo {year} {2001})}\BibitemShut {NoStop}%
\bibitem [{\citenamefont {Krausz}\ and\ \citenamefont
  {Ivanov}(2009)}]{krausz2009a}%
  \BibitemOpen
  \bibfield  {author} {\bibinfo {author} {\bibfnamefont {F.}~\bibnamefont
  {Krausz}}\ and\ \bibinfo {author} {\bibfnamefont {M.}~\bibnamefont
  {Ivanov}},\ }\bibfield  {title} {\enquote {\bibinfo {title} {Attosecond
  physics},}\ }\href {\doibase 10.1103/RevModPhys.81.163} {\bibfield  {journal}
  {\bibinfo  {journal} {Rev. Mod. Phys.}\ }\textbf {\bibinfo {volume} {81}},\
  \bibinfo {pages} {163} (\bibinfo {year} {2009})}\BibitemShut {NoStop}%
\bibitem [{\citenamefont {Sali{\`e}res}\ \emph {et~al.}(2012)\citenamefont
  {Sali{\`e}res}, \citenamefont {Maquet}, \citenamefont {Haessler},
  \citenamefont {Caillat},\ and\ \citenamefont {Ta{\"\i}eb}}]{salieres2012a}%
  \BibitemOpen
  \bibfield  {author} {\bibinfo {author} {\bibfnamefont {P.}~\bibnamefont
  {Sali{\`e}res}}, \bibinfo {author} {\bibfnamefont {A.}~\bibnamefont
  {Maquet}}, \bibinfo {author} {\bibfnamefont {S.}~\bibnamefont {Haessler}},
  \bibinfo {author} {\bibfnamefont {J.}~\bibnamefont {Caillat}}, \ and\
  \bibinfo {author} {\bibfnamefont {R.}~\bibnamefont {Ta{\"\i}eb}},\ }\bibfield
   {title} {\enquote {\bibinfo {title} {{Imaging orbitals with attosecond and
  {\AA}ngstr{\"o}m resolutions: toward attochemistry?}}}\ }\href
  {http://stacks.iop.org/0034-4885/75/i=6/a=062401} {\bibfield  {journal}
  {\bibinfo  {journal} {Rep. Prog. Phys.}\ }\textbf {\bibinfo {volume} {75}},\
  \bibinfo {pages} {062401} (\bibinfo {year} {2012})}\BibitemShut {NoStop}%
\bibitem [{\citenamefont {Merritt}\ \emph {et~al.}(2021)\citenamefont
  {Merritt}, \citenamefont {Jacquemin},\ and\ \citenamefont
  {Vacher}}]{merritt2021a}%
  \BibitemOpen
  \bibfield  {author} {\bibinfo {author} {\bibfnamefont {I.~C.~D.}\
  \bibnamefont {Merritt}}, \bibinfo {author} {\bibfnamefont {D.}~\bibnamefont
  {Jacquemin}}, \ and\ \bibinfo {author} {\bibfnamefont {M.}~\bibnamefont
  {Vacher}},\ }\bibfield  {title} {\enquote {\bibinfo {title} {Attochemistry:
  Is controlling electrons the future of photochemistry?}}\ } {}
  {\bibfield  {journal} {\bibinfo  {journal} {J. Phys. Chem. Lett.}\ }\textbf
  {\bibinfo {volume} {12}},\ \bibinfo {pages} {8404} (\bibinfo {year}
  {2021})}\BibitemShut {NoStop}%
\bibitem [{\citenamefont {Calegari}\ and\ \citenamefont
  {Mart{\'\i}n}(2023)}]{calegari2023a}%
  \BibitemOpen
  \bibfield  {author} {\bibinfo {author} {\bibfnamefont {F.}~\bibnamefont
  {Calegari}}\ and\ \bibinfo {author} {\bibfnamefont {F.}~\bibnamefont
  {Mart{\'\i}n}},\ }\bibfield  {title} {\enquote {\bibinfo {title} {Open
  questions in attochemistry},}\ }\href {\doibase
  https://doi.org/10.1038/s42004-023-00989-0} {\bibfield  {journal} {\bibinfo
  {journal} {Commun Chem}\ }\textbf {\bibinfo {volume} {6}},\ \bibinfo {pages}
  {184} (\bibinfo {year} {2023})}\BibitemShut {NoStop}%
\bibitem [{\citenamefont {Barillot}\ \emph {et~al.}(2015)\citenamefont
  {Barillot}, \citenamefont {Cauchy}, \citenamefont {Hervieux}, \citenamefont
  {Gisselbrecht}, \citenamefont {Canton}, \citenamefont {Johnsson},
  \citenamefont {Laksman}, \citenamefont {Mansson}, \citenamefont
  {Dahlstr\"om}, \citenamefont {Magrakvelidze}, \citenamefont {Dixit},
  \citenamefont {Madjet}, \citenamefont {Chakraborty}, \citenamefont {Suraud},
  \citenamefont {Dinh}, \citenamefont {Wopperer}, \citenamefont {Hansen},
  \citenamefont {Loriot}, \citenamefont {Bordas}, \citenamefont {Sorensen},\
  and\ \citenamefont {L\'epine}}]{barillot2015a}%
  \BibitemOpen
  \bibfield  {author} {\bibinfo {author} {\bibfnamefont {T.}~\bibnamefont
  {Barillot}}, \bibinfo {author} {\bibfnamefont {C.}~\bibnamefont {Cauchy}},
  \bibinfo {author} {\bibfnamefont {P.-A.}\ \bibnamefont {Hervieux}}, \bibinfo
  {author} {\bibfnamefont {M.}~\bibnamefont {Gisselbrecht}}, \bibinfo {author}
  {\bibfnamefont {S.~E.}\ \bibnamefont {Canton}}, \bibinfo {author}
  {\bibfnamefont {P.}~\bibnamefont {Johnsson}}, \bibinfo {author}
  {\bibfnamefont {J.}~\bibnamefont {Laksman}}, \bibinfo {author} {\bibfnamefont
  {E.~P.}\ \bibnamefont {Mansson}}, \bibinfo {author} {\bibfnamefont {J.~M.}\
  \bibnamefont {Dahlstr\"om}}, \bibinfo {author} {\bibfnamefont
  {M.}~\bibnamefont {Magrakvelidze}}, \bibinfo {author} {\bibfnamefont
  {G.}~\bibnamefont {Dixit}}, \bibinfo {author} {\bibfnamefont {M.~E.}\
  \bibnamefont {Madjet}}, \bibinfo {author} {\bibfnamefont {H.~S.}\
  \bibnamefont {Chakraborty}}, \bibinfo {author} {\bibfnamefont
  {E.}~\bibnamefont {Suraud}}, \bibinfo {author} {\bibfnamefont {P.~M.}\
  \bibnamefont {Dinh}}, \bibinfo {author} {\bibfnamefont {P.}~\bibnamefont
  {Wopperer}}, \bibinfo {author} {\bibfnamefont {K.}~\bibnamefont {Hansen}},
  \bibinfo {author} {\bibfnamefont {V.}~\bibnamefont {Loriot}}, \bibinfo
  {author} {\bibfnamefont {C.}~\bibnamefont {Bordas}}, \bibinfo {author}
  {\bibfnamefont {S.}~\bibnamefont {Sorensen}}, \ and\ \bibinfo {author}
  {\bibfnamefont {F.}~\bibnamefont {L\'epine}},\ }\bibfield  {title} {\enquote
  {\bibinfo {title} {{Angular asymmetry and attosecond time delay from the
  giant plasmon resonance in C$_{60}$ photoionization}},}\ }\href {\doibase
  10.1103/PhysRevA.91.033413} {\bibfield  {journal} {\bibinfo  {journal} {Phys.
  Rev. A}\ }\textbf {\bibinfo {volume} {91}},\ \bibinfo {pages} {033413}
  (\bibinfo {year} {2015})}\BibitemShut {NoStop}%
\bibitem [{\citenamefont {Huppert}\ \emph {et~al.}(2016)\citenamefont
  {Huppert}, \citenamefont {Jordan}, \citenamefont {Baykusheva}, \citenamefont
  {von Conta},\ and\ \citenamefont {W\"orner}}]{huppert2016a}%
  \BibitemOpen
  \bibfield  {author} {\bibinfo {author} {\bibfnamefont {M.}~\bibnamefont
  {Huppert}}, \bibinfo {author} {\bibfnamefont {I.}~\bibnamefont {Jordan}},
  \bibinfo {author} {\bibfnamefont {D.}~\bibnamefont {Baykusheva}}, \bibinfo
  {author} {\bibfnamefont {A.}~\bibnamefont {von Conta}}, \ and\ \bibinfo
  {author} {\bibfnamefont {H.~J.}\ \bibnamefont {W\"orner}},\ }\bibfield
  {title} {\enquote {\bibinfo {title} {Attosecond delays in molecular
  photoionization},}\ }\href {\doibase 10.1103/PhysRevLett.117.093001}
  {\bibfield  {journal} {\bibinfo  {journal} {Phys. Rev. Lett.}\ }\textbf
  {\bibinfo {volume} {117}},\ \bibinfo {pages} {093001} (\bibinfo {year}
  {2016})}\BibitemShut {NoStop}%
\bibitem [{\citenamefont {Bray}\ \emph
  {et~al.}(2018{\natexlab{a}})\citenamefont {Bray}, \citenamefont {Naseem},\
  and\ \citenamefont {Kheifets}}]{bray2018b}%
  \BibitemOpen
  \bibfield  {author} {\bibinfo {author} {\bibfnamefont {A.~W.}\ \bibnamefont
  {Bray}}, \bibinfo {author} {\bibfnamefont {F.}~\bibnamefont {Naseem}}, \ and\
  \bibinfo {author} {\bibfnamefont {A.~S.}\ \bibnamefont {Kheifets}},\
  }\bibfield  {title} {\enquote {\bibinfo {title} {Photoionization of {Xe} and
  {Xe}$@${C}$_{60}$ from the $4d$ shell in {RABBITT} fields},}\ }\href
  {\doibase 10.1103/PhysRevA.98.043427} {\bibfield  {journal} {\bibinfo
  {journal} {Phys. Rev. A}\ }\textbf {\bibinfo {volume} {98}},\ \bibinfo
  {pages} {043427} (\bibinfo {year} {2018}{\natexlab{a}})}\BibitemShut
  {NoStop}%
\bibitem [{\citenamefont {Ahmadi}\ \emph {et~al.}(2022)\citenamefont {Ahmadi},
  \citenamefont {Pl{\'e}siat}, \citenamefont {Moioli}, \citenamefont
  {Frassetto}, \citenamefont {Poletto}, \citenamefont {Decleva}, \citenamefont
  {Schr{\"o}ter}, \citenamefont {Pfeifer}, \citenamefont {Moshammer},
  \citenamefont {Palacios}, \citenamefont {Mart{\'\i}n},\ and\ \citenamefont
  {Sansone}}]{ahmadi2022a}%
  \BibitemOpen
  \bibfield  {author} {\bibinfo {author} {\bibfnamefont {H.}~\bibnamefont
  {Ahmadi}}, \bibinfo {author} {\bibfnamefont {E.}~\bibnamefont {Pl{\'e}siat}},
  \bibinfo {author} {\bibfnamefont {M.}~\bibnamefont {Moioli}}, \bibinfo
  {author} {\bibfnamefont {F.}~\bibnamefont {Frassetto}}, \bibinfo {author}
  {\bibfnamefont {L.}~\bibnamefont {Poletto}}, \bibinfo {author} {\bibfnamefont
  {P.}~\bibnamefont {Decleva}}, \bibinfo {author} {\bibfnamefont {C.~D.}\
  \bibnamefont {Schr{\"o}ter}}, \bibinfo {author} {\bibfnamefont
  {T.}~\bibnamefont {Pfeifer}}, \bibinfo {author} {\bibfnamefont
  {R.}~\bibnamefont {Moshammer}}, \bibinfo {author} {\bibfnamefont
  {A.}~\bibnamefont {Palacios}}, \bibinfo {author} {\bibfnamefont
  {F.}~\bibnamefont {Mart{\'\i}n}}, \ and\ \bibinfo {author} {\bibfnamefont
  {G.}~\bibnamefont {Sansone}},\ }\bibfield  {title} {\enquote {\bibinfo
  {title} {{Attosecond photoionisation time delays reveal the anisotropy of the
  molecular potential in the recoil frame}},}\ } {} {\bibfield
  {journal} {\bibinfo  {journal} {Nat Commun}\ }\textbf {\bibinfo {volume}
  {{13}}},\ \bibinfo {pages} {1242} (\bibinfo {year} {2022})}\BibitemShut
  {NoStop}%
\bibitem [{\citenamefont {Boyer}\ \emph {et~al.}(2023)\citenamefont {Boyer},
  \citenamefont {Nandi},\ and\ \citenamefont {Loriot}}]{boyer2023a}%
  \BibitemOpen
  \bibfield  {author} {\bibinfo {author} {\bibfnamefont {A.}~\bibnamefont
  {Boyer}}, \bibinfo {author} {\bibfnamefont {S.}~\bibnamefont {Nandi}}, \ and\
  \bibinfo {author} {\bibfnamefont {V.}~\bibnamefont {Loriot}},\ }\bibfield
  {title} {\enquote {\bibinfo {title} {Attosecond probing of photoionization
  dynamics from diatomic to many-atom molecules},}\ }\href {\doibase
  10.1140/epjs/s11734-022-00754-9} {\bibfield  {journal} {\bibinfo  {journal}
  {Eur. Phys. J. Spec. Top.}\ } (\bibinfo {year} {2023}),\
  10.1140/epjs/s11734-022-00754-9}\BibitemShut {NoStop}%
\bibitem [{\citenamefont {Gong}\ \emph {et~al.}(2023)\citenamefont {Gong},
  \citenamefont {Pl\'esiat}, \citenamefont {Palacios}, \citenamefont {Heck},
  \citenamefont {Mart{\`\i}n},\ and\ \citenamefont {W{\"o}rner}}]{gong2023a}%
  \BibitemOpen
  \bibfield  {author} {\bibinfo {author} {\bibfnamefont {X.}~\bibnamefont
  {Gong}}, \bibinfo {author} {\bibfnamefont {E.}~\bibnamefont {Pl\'esiat}},
  \bibinfo {author} {\bibfnamefont {A.}~\bibnamefont {Palacios}}, \bibinfo
  {author} {\bibfnamefont {S.}~\bibnamefont {Heck}}, \bibinfo {author}
  {\bibfnamefont {F.}~\bibnamefont {Mart{\`\i}n}}, \ and\ \bibinfo {author}
  {\bibfnamefont {H.~J.}\ \bibnamefont {W{\"o}rner}},\ }\bibfield  {title}
  {\enquote {\bibinfo {title} {{Attosecond delays between dissociative and
  non-dissociative ionization of polyatomic molecules}},}\ } {}
  {\bibfield  {journal} {\bibinfo  {journal} {Nat Commun}\ }\textbf {\bibinfo
  {volume} {14}},\ \bibinfo {pages} {4402} (\bibinfo {year}
  {2023})}\BibitemShut {NoStop}%
\bibitem [{\citenamefont {Thuppilakkadan}\ \emph {et~al.}(2023)\citenamefont
  {Thuppilakkadan}, \citenamefont {Banerjee},\ and\ \citenamefont
  {Varma}}]{thuppilakkadan2023a}%
  \BibitemOpen
  \bibfield  {author} {\bibinfo {author} {\bibfnamefont {A.}~\bibnamefont
  {Thuppilakkadan}}, \bibinfo {author} {\bibfnamefont {S.}~\bibnamefont
  {Banerjee}}, \ and\ \bibinfo {author} {\bibfnamefont {H.~R.}\ \bibnamefont
  {Varma}},\ }\bibfield  {title} {\enquote {\bibinfo {title} {Modifications in
  the angular photoemission time delay in
  $\mathrm{Ar}@{\mathrm{c}}_{60}^{q=\ensuremath{-}1}$: Coulomb confinement
  resonance as an amplifier of the spin-orbit-interaction-activated
  interchannel coupling effect},}\ }\href {\doibase
  10.1103/PhysRevA.107.052804} {\bibfield  {journal} {\bibinfo  {journal}
  {Phys. Rev. A}\ }\textbf {\bibinfo {volume} {107}},\ \bibinfo {pages}
  {052804} (\bibinfo {year} {2023})}\BibitemShut {NoStop}%
\bibitem [{\citenamefont {Beaulieu}\ \emph {et~al.}(2017)\citenamefont
  {Beaulieu}, \citenamefont {Comby}, \citenamefont {Clergerie}, \citenamefont
  {Caillat}, \citenamefont {Descamps}, \citenamefont {Dudovich}, \citenamefont
  {Fabre}, \citenamefont {G{\'e}neaux}, \citenamefont {L{\'e}gar{\'e}},
  \citenamefont {Petit}, \citenamefont {Pons}, \citenamefont {Porat},
  \citenamefont {Ruchon}, \citenamefont {Ta{\"\i}eb}, \citenamefont
  {Blanchet},\ and\ \citenamefont {Mairesse}}]{beaulieu2017a}%
  \BibitemOpen
  \bibfield  {author} {\bibinfo {author} {\bibfnamefont {S.}~\bibnamefont
  {Beaulieu}}, \bibinfo {author} {\bibfnamefont {A.}~\bibnamefont {Comby}},
  \bibinfo {author} {\bibfnamefont {A.}~\bibnamefont {Clergerie}}, \bibinfo
  {author} {\bibfnamefont {J.}~\bibnamefont {Caillat}}, \bibinfo {author}
  {\bibfnamefont {D.}~\bibnamefont {Descamps}}, \bibinfo {author}
  {\bibfnamefont {N.}~\bibnamefont {Dudovich}}, \bibinfo {author}
  {\bibfnamefont {B.}~\bibnamefont {Fabre}}, \bibinfo {author} {\bibfnamefont
  {R.}~\bibnamefont {G{\'e}neaux}}, \bibinfo {author} {\bibfnamefont
  {F.}~\bibnamefont {L{\'e}gar{\'e}}}, \bibinfo {author} {\bibfnamefont
  {S.}~\bibnamefont {Petit}}, \bibinfo {author} {\bibfnamefont
  {B.}~\bibnamefont {Pons}}, \bibinfo {author} {\bibfnamefont {G.}~\bibnamefont
  {Porat}}, \bibinfo {author} {\bibfnamefont {T.}~\bibnamefont {Ruchon}},
  \bibinfo {author} {\bibfnamefont {R.}~\bibnamefont {Ta{\"\i}eb}}, \bibinfo
  {author} {\bibfnamefont {V.}~\bibnamefont {Blanchet}}, \ and\ \bibinfo
  {author} {\bibfnamefont {Y.}~\bibnamefont {Mairesse}},\ }\bibfield  {title}
  {\enquote {\bibinfo {title} {Attosecond-resolved photoionization of chiral
  molecules},}\ } {} {\bibfield  {journal} {\bibinfo  {journal}
  {Science}\ }\textbf {\bibinfo {volume} {358}},\ \bibinfo {pages} {1288}
  (\bibinfo {year} {2017})}\BibitemShut {NoStop}%
\bibitem [{\citenamefont {Hockett}\ \emph {et~al.}(2016)\citenamefont
  {Hockett}, \citenamefont {Frumker}, \citenamefont {Villeneuve},\ and\
  \citenamefont {Corkum}}]{hockett2016a}%
  \BibitemOpen
  \bibfield  {author} {\bibinfo {author} {\bibfnamefont {P.}~\bibnamefont
  {Hockett}}, \bibinfo {author} {\bibfnamefont {E.}~\bibnamefont {Frumker}},
  \bibinfo {author} {\bibfnamefont {D.~M.}\ \bibnamefont {Villeneuve}}, \ and\
  \bibinfo {author} {\bibfnamefont {P.~B.}\ \bibnamefont {Corkum}},\ }\bibfield
   {title} {\enquote {\bibinfo {title} {Time delay in molecular
  photoionization},}\ }\href {\doibase 10.1088/0953-4075/49/9/095602}
  {\bibfield  {journal} {\bibinfo  {journal} {J. Phys. B: At. Mol. Opt. Phys.}\
  }\textbf {\bibinfo {volume} {49}},\ \bibinfo {pages} {095602} (\bibinfo
  {year} {2016})}\BibitemShut {NoStop}%
\bibitem [{\citenamefont {Zhang}\ \emph {et~al.}(2023)\citenamefont {Zhang},
  \citenamefont {Wang}, \citenamefont {Liao},\ and\ \citenamefont
  {Lu}}]{zhang2023a}%
  \BibitemOpen
  \bibfield  {author} {\bibinfo {author} {\bibfnamefont {R.}~\bibnamefont
  {Zhang}}, \bibinfo {author} {\bibfnamefont {F.}~\bibnamefont {Wang}},
  \bibinfo {author} {\bibfnamefont {Q.}~\bibnamefont {Liao}}, \ and\ \bibinfo
  {author} {\bibfnamefont {P.}~\bibnamefont {Lu}},\ }\bibfield  {title}
  {\enquote {\bibinfo {title} {{Role of molecular alignment in attosecond
  photoionization of ${\mathrm{N}}_{2}$}},}\ }\href {\doibase
  10.1103/PhysRevA.108.013113} {\bibfield  {journal} {\bibinfo  {journal}
  {Phys. Rev. A}\ }\textbf {\bibinfo {volume} {108}},\ \bibinfo {pages}
  {013113} (\bibinfo {year} {2023})}\BibitemShut {NoStop}%
\bibitem [{\citenamefont {Ke}\ \emph {et~al.}(2023)\citenamefont {Ke},
  \citenamefont {Zhou}, \citenamefont {Liao}, \citenamefont {Li}, \citenamefont
  {Liu},\ and\ \citenamefont {Lu}}]{ke2023a}%
  \BibitemOpen
  \bibfield  {author} {\bibinfo {author} {\bibfnamefont {Q.}~\bibnamefont
  {Ke}}, \bibinfo {author} {\bibfnamefont {Y.}~\bibnamefont {Zhou}}, \bibinfo
  {author} {\bibfnamefont {Y.}~\bibnamefont {Liao}}, \bibinfo {author}
  {\bibfnamefont {M.}~\bibnamefont {Li}}, \bibinfo {author} {\bibfnamefont
  {K.}~\bibnamefont {Liu}}, \ and\ \bibinfo {author} {\bibfnamefont
  {P.}~\bibnamefont {Lu}},\ }\bibfield  {title} {\enquote {\bibinfo {title}
  {{Spheroidal-wave analysis of time delay in molecular reconstruction of
  attosecond beating by interference of two-photon transitions around a
  Cooper-like minimum}},}\ }\href {\doibase 10.1103/PhysRevA.108.013112}
  {\bibfield  {journal} {\bibinfo  {journal} {Phys. Rev. A}\ }\textbf {\bibinfo
  {volume} {108}},\ \bibinfo {pages} {013112} (\bibinfo {year}
  {2023})}\BibitemShut {NoStop}%
\bibitem [{\citenamefont {Chacon}\ \emph {et~al.}(2014)\citenamefont {Chacon},
  \citenamefont {Lein},\ and\ \citenamefont {Ruiz}}]{chacon2014a}%
  \BibitemOpen
  \bibfield  {author} {\bibinfo {author} {\bibfnamefont {A.}~\bibnamefont
  {Chacon}}, \bibinfo {author} {\bibfnamefont {M.}~\bibnamefont {Lein}}, \ and\
  \bibinfo {author} {\bibfnamefont {C.}~\bibnamefont {Ruiz}},\ }\bibfield
  {title} {\enquote {\bibinfo {title} {{Asymmetry of Wigner's time delay in a
  small molecule}},}\ }\href {\doibase 10.1103/PhysRevA.89.053427} {\bibfield
  {journal} {\bibinfo  {journal} {Phys. Rev. A}\ }\textbf {\bibinfo {volume}
  {89}},\ \bibinfo {pages} {053427} (\bibinfo {year} {2014})}\BibitemShut
  {NoStop}%
\bibitem [{\citenamefont {Heuser}\ \emph {et~al.}(2016)\citenamefont {Heuser},
  \citenamefont {Jim\'enez~Gal\'an}, \citenamefont {Cirelli}, \citenamefont
  {Marante}, \citenamefont {Sabbar}, \citenamefont {Boge}, \citenamefont
  {Lucchini}, \citenamefont {Gallmann}, \citenamefont {Ivanov}, \citenamefont
  {Kheifets}, \citenamefont {Dahlstr\"om}, \citenamefont {Lindroth},
  \citenamefont {Argenti}, \citenamefont {Mart\'{\i}n},\ and\ \citenamefont
  {Keller}}]{heuser2016a}%
  \BibitemOpen
  \bibfield  {author} {\bibinfo {author} {\bibfnamefont {S.}~\bibnamefont
  {Heuser}}, \bibinfo {author} {\bibfnamefont {A.}~\bibnamefont
  {Jim\'enez~Gal\'an}}, \bibinfo {author} {\bibfnamefont {C.}~\bibnamefont
  {Cirelli}}, \bibinfo {author} {\bibfnamefont {C.}~\bibnamefont {Marante}},
  \bibinfo {author} {\bibfnamefont {M.}~\bibnamefont {Sabbar}}, \bibinfo
  {author} {\bibfnamefont {R.}~\bibnamefont {Boge}}, \bibinfo {author}
  {\bibfnamefont {M.}~\bibnamefont {Lucchini}}, \bibinfo {author}
  {\bibfnamefont {L.}~\bibnamefont {Gallmann}}, \bibinfo {author}
  {\bibfnamefont {I.}~\bibnamefont {Ivanov}}, \bibinfo {author} {\bibfnamefont
  {A.~S.}\ \bibnamefont {Kheifets}}, \bibinfo {author} {\bibfnamefont {J.~M.}\
  \bibnamefont {Dahlstr\"om}}, \bibinfo {author} {\bibfnamefont
  {E.}~\bibnamefont {Lindroth}}, \bibinfo {author} {\bibfnamefont
  {L.}~\bibnamefont {Argenti}}, \bibinfo {author} {\bibfnamefont
  {F.}~\bibnamefont {Mart\'{\i}n}}, \ and\ \bibinfo {author} {\bibfnamefont
  {U.}~\bibnamefont {Keller}},\ }\bibfield  {title} {\enquote {\bibinfo {title}
  {{Angular dependence of photoemission time delay in helium}},}\ }\href
  {\doibase 10.1103/PhysRevA.94.063409} {\bibfield  {journal} {\bibinfo
  {journal} {Phys. Rev. A}\ }\textbf {\bibinfo {volume} {94}},\ \bibinfo
  {pages} {063409} (\bibinfo {year} {2016})}\BibitemShut {NoStop}%
\bibitem [{\citenamefont {Hockett}(2017)}]{hockett2017a}%
  \BibitemOpen
  \bibfield  {author} {\bibinfo {author} {\bibfnamefont {P.}~\bibnamefont
  {Hockett}},\ }\bibfield  {title} {\enquote {\bibinfo {title} {Angle-resolved
  {RABBITT}: theory and numerics},}\ }\href {\doibase 10.1088/1361-6455/aa7887}
  {\bibfield  {journal} {\bibinfo  {journal} {J. Phys. B: At. Mol. Opt. Phys.}\
  }\textbf {\bibinfo {volume} {50}},\ \bibinfo {pages} {154002} (\bibinfo
  {year} {2017})}\BibitemShut {NoStop}%
\bibitem [{\citenamefont {Bray}\ \emph
  {et~al.}(2018{\natexlab{b}})\citenamefont {Bray}, \citenamefont {Naseem},\
  and\ \citenamefont {Kheifets}}]{bray2018a}%
  \BibitemOpen
  \bibfield  {author} {\bibinfo {author} {\bibfnamefont {A.~W.}\ \bibnamefont
  {Bray}}, \bibinfo {author} {\bibfnamefont {F.}~\bibnamefont {Naseem}}, \ and\
  \bibinfo {author} {\bibfnamefont {A.~S.}\ \bibnamefont {Kheifets}},\
  }\bibfield  {title} {\enquote {\bibinfo {title} {Simulation of
  angular-resolved rabbitt measurements in noble-gas atoms},}\ }\href {\doibase
  10.1103/PhysRevA.97.063404} {\bibfield  {journal} {\bibinfo  {journal} {Phys.
  Rev. A}\ }\textbf {\bibinfo {volume} {97}},\ \bibinfo {pages} {063404}
  (\bibinfo {year} {2018}{\natexlab{b}})}\BibitemShut {NoStop}%
\bibitem [{\citenamefont {Fuchs}\ \emph {et~al.}(2020)\citenamefont {Fuchs},
  \citenamefont {Douguet}, \citenamefont {Donsa}, \citenamefont {Martin},
  \citenamefont {Burgd\"{o}rfer}, \citenamefont {Argenti}, \citenamefont
  {Cattaneo},\ and\ \citenamefont {Keller}}]{fuchs2020a}%
  \BibitemOpen
  \bibfield  {author} {\bibinfo {author} {\bibfnamefont {J.}~\bibnamefont
  {Fuchs}}, \bibinfo {author} {\bibfnamefont {N.}~\bibnamefont {Douguet}},
  \bibinfo {author} {\bibfnamefont {S.}~\bibnamefont {Donsa}}, \bibinfo
  {author} {\bibfnamefont {F.}~\bibnamefont {Martin}}, \bibinfo {author}
  {\bibfnamefont {J.}~\bibnamefont {Burgd\"{o}rfer}}, \bibinfo {author}
  {\bibfnamefont {L.}~\bibnamefont {Argenti}}, \bibinfo {author} {\bibfnamefont
  {L.}~\bibnamefont {Cattaneo}}, \ and\ \bibinfo {author} {\bibfnamefont
  {U.}~\bibnamefont {Keller}},\ }\bibfield  {title} {\enquote {\bibinfo {title}
  {Time delays from one-photon transitions in the continuum},}\ }\href
  {\doibase 10.1364/OPTICA.378639} {\bibfield  {journal} {\bibinfo  {journal}
  {Optica}\ }\textbf {\bibinfo {volume} {7}},\ \bibinfo {pages} {154} (\bibinfo
  {year} {2020})}\BibitemShut {NoStop}%
\bibitem [{cir()}]{cirelli2018a}%
  \BibitemOpen
  \bibfield  {title} {\enquote {\bibinfo {title} {Anisotropic photoemission
  time delays close to a fano resonance},}\ } {} {\ }\BibitemShut
  {NoStop}%
\bibitem [{\citenamefont {Trabert}\ \emph {et~al.}(2023)\citenamefont
  {Trabert}, \citenamefont {Anders}, \citenamefont {Geyer}, \citenamefont
  {Hofmann}, \citenamefont {Sch\"offler}, \citenamefont {Schmidt},
  \citenamefont {Jahnke}, \citenamefont {Kunitski}, \citenamefont {D\"orner},\
  and\ \citenamefont {Eckart}}]{trabert2023a}%
  \BibitemOpen
  \bibfield  {author} {\bibinfo {author} {\bibfnamefont {D.}~\bibnamefont
  {Trabert}}, \bibinfo {author} {\bibfnamefont {N.}~\bibnamefont {Anders}},
  \bibinfo {author} {\bibfnamefont {A.}~\bibnamefont {Geyer}}, \bibinfo
  {author} {\bibfnamefont {M.}~\bibnamefont {Hofmann}}, \bibinfo {author}
  {\bibfnamefont {M.~S.}\ \bibnamefont {Sch\"offler}}, \bibinfo {author}
  {\bibfnamefont {L.~P.~H.}\ \bibnamefont {Schmidt}}, \bibinfo {author}
  {\bibfnamefont {T.}~\bibnamefont {Jahnke}}, \bibinfo {author} {\bibfnamefont
  {M.}~\bibnamefont {Kunitski}}, \bibinfo {author} {\bibfnamefont
  {R.}~\bibnamefont {D\"orner}}, \ and\ \bibinfo {author} {\bibfnamefont
  {S.}~\bibnamefont {Eckart}},\ }\bibfield  {title} {\enquote {\bibinfo {title}
  {{Angular dependence of the Wigner time delay upon strong-field ionization
  from an aligned $p$ orbital}},}\ }\href {\doibase
  10.1103/PhysRevResearch.5.023118} {\bibfield  {journal} {\bibinfo  {journal}
  {Phys. Rev. Res.}\ }\textbf {\bibinfo {volume} {5}},\ \bibinfo {pages}
  {023118} (\bibinfo {year} {2023})}\BibitemShut {NoStop}%
\bibitem [{\citenamefont {Vos}\ \emph {et~al.}(2018)\citenamefont {Vos},
  \citenamefont {Cattaneo}, \citenamefont {Patchkovskii}, \citenamefont
  {Zimmermann}, \citenamefont {Cirelli}, \citenamefont {Lucchini},
  \citenamefont {Kheifets}, \citenamefont {Landsman},\ and\ \citenamefont
  {Keller}}]{vos2018a}%
  \BibitemOpen
  \bibfield  {author} {\bibinfo {author} {\bibfnamefont {J.}~\bibnamefont
  {Vos}}, \bibinfo {author} {\bibfnamefont {L.}~\bibnamefont {Cattaneo}},
  \bibinfo {author} {\bibfnamefont {S.}~\bibnamefont {Patchkovskii}}, \bibinfo
  {author} {\bibfnamefont {T.}~\bibnamefont {Zimmermann}}, \bibinfo {author}
  {\bibfnamefont {C.}~\bibnamefont {Cirelli}}, \bibinfo {author} {\bibfnamefont
  {M.}~\bibnamefont {Lucchini}}, \bibinfo {author} {\bibfnamefont
  {A.}~\bibnamefont {Kheifets}}, \bibinfo {author} {\bibfnamefont {A.~S.}\
  \bibnamefont {Landsman}}, \ and\ \bibinfo {author} {\bibfnamefont
  {U.}~\bibnamefont {Keller}},\ }\bibfield  {title} {\enquote {\bibinfo {title}
  {{Orientation-dependent stereo Wigner time delay and electron localization in
  a small molecule}},}\ } {} {\bibfield  {journal} {\bibinfo
  {journal} {Science}\ }\textbf {\bibinfo {volume} {360}},\ \bibinfo {pages}
  {1326} (\bibinfo {year} {2018})}\BibitemShut {NoStop}%
\bibitem [{\citenamefont {Desrier}(2018)}]{desrierphdthesis}%
  \BibitemOpen
  \bibfield  {author} {\bibinfo {author} {\bibfnamefont {A.}~\bibnamefont
  {Desrier}},\ }\emph {\bibinfo {title} {{Dynamique ultrarapide corr\'el\'ee :
  th\'eorie, simulations et interpr\'etations d'exp\'eriences de spectroscopie
  ``attoseconde''}}},\  {} {Ph.D. thesis},\ \bibinfo  {school}
  {{Sorbonne Universit\'e}} (\bibinfo {year} {{2018}})\BibitemShut {NoStop}%
\bibitem [{\citenamefont {Gong}\ \emph {et~al.}(2022)\citenamefont {Gong},
  \citenamefont {Jiang}, \citenamefont {Tong}, \citenamefont {Qiang},
  \citenamefont {Lu}, \citenamefont {Ni}, \citenamefont {Lucchese},
  \citenamefont {Ueda},\ and\ \citenamefont {Wu}}]{gong2022a}%
  \BibitemOpen
  \bibfield  {author} {\bibinfo {author} {\bibfnamefont {X.}~\bibnamefont
  {Gong}}, \bibinfo {author} {\bibfnamefont {W.}~\bibnamefont {Jiang}},
  \bibinfo {author} {\bibfnamefont {J.}~\bibnamefont {Tong}}, \bibinfo {author}
  {\bibfnamefont {J.}~\bibnamefont {Qiang}}, \bibinfo {author} {\bibfnamefont
  {P.}~\bibnamefont {Lu}}, \bibinfo {author} {\bibfnamefont {H.}~\bibnamefont
  {Ni}}, \bibinfo {author} {\bibfnamefont {R.}~\bibnamefont {Lucchese}},
  \bibinfo {author} {\bibfnamefont {K.}~\bibnamefont {Ueda}}, \ and\ \bibinfo
  {author} {\bibfnamefont {J.}~\bibnamefont {Wu}},\ }\bibfield  {title}
  {\enquote {\bibinfo {title} {Asymmetric attosecond photoionization in
  molecular shape resonance},}\ } {} {\bibfield  {journal} {\bibinfo
  {journal} {Phys. Rev. X}\ }\textbf {\bibinfo {volume} {12}},\ \bibinfo
  {pages} {011002} (\bibinfo {year} {2022})}\BibitemShut {NoStop}%
\bibitem [{\citenamefont {Liao}\ \emph {et~al.}(2021)\citenamefont {Liao},
  \citenamefont {Zhou}, \citenamefont {Pi}, \citenamefont {Ke}, \citenamefont
  {Liang}, \citenamefont {Zhao}, \citenamefont {Li},\ and\ \citenamefont
  {Lu}}]{liao2021a}%
  \BibitemOpen
  \bibfield  {author} {\bibinfo {author} {\bibfnamefont {Y.}~\bibnamefont
  {Liao}}, \bibinfo {author} {\bibfnamefont {Y.}~\bibnamefont {Zhou}}, \bibinfo
  {author} {\bibfnamefont {L.-W.}\ \bibnamefont {Pi}}, \bibinfo {author}
  {\bibfnamefont {Q.}~\bibnamefont {Ke}}, \bibinfo {author} {\bibfnamefont
  {J.}~\bibnamefont {Liang}}, \bibinfo {author} {\bibfnamefont
  {Y.}~\bibnamefont {Zhao}}, \bibinfo {author} {\bibfnamefont {M.}~\bibnamefont
  {Li}}, \ and\ \bibinfo {author} {\bibfnamefont {P.}~\bibnamefont {Lu}},\
  }\bibfield  {title} {\enquote {\bibinfo {title} {{Two-center interference and
  stereo Wigner time delay in photoionization of asymmetric molecules}},}\
  }\href {\doibase 10.1103/PhysRevA.104.013110} {\bibfield  {journal} {\bibinfo
   {journal} {Phys. Rev. A}\ }\textbf {\bibinfo {volume} {104}},\ \bibinfo
  {pages} {013110} (\bibinfo {year} {2021})}\BibitemShut {NoStop}%
\bibitem [{\citenamefont {Maquet}\ \emph {et~al.}(2014)\citenamefont {Maquet},
  \citenamefont {Caillat},\ and\ \citenamefont {Ta{\"\i}eb}}]{maquet2014a}%
  \BibitemOpen
  \bibfield  {author} {\bibinfo {author} {\bibfnamefont {A.}~\bibnamefont
  {Maquet}}, \bibinfo {author} {\bibfnamefont {J.}~\bibnamefont {Caillat}}, \
  and\ \bibinfo {author} {\bibfnamefont {R.}~\bibnamefont {Ta{\"\i}eb}},\
  }\bibfield  {title} {\enquote {\bibinfo {title} {Attosecond delays in
  photoionization: time {\em and} quantum mechanics},}\ }\href
  {http://stacks.iop.org/0953-4075/47/i=20/a=204004} {\bibfield  {journal}
  {\bibinfo  {journal} {J. Phys. B: At. Mol. Opt. Phys.}\ }\textbf {\bibinfo
  {volume} {47}},\ \bibinfo {pages} {204004} (\bibinfo {year}
  {2014})}\BibitemShut {NoStop}%
\bibitem [{\citenamefont {Dahlstr{\"o}m}\ \emph {et~al.}(2012)\citenamefont
  {Dahlstr{\"o}m}, \citenamefont {L'Huillier},\ and\ \citenamefont
  {Maquet}}]{dahlstrom2012a}%
  \BibitemOpen
  \bibfield  {author} {\bibinfo {author} {\bibfnamefont {J.~M.}\ \bibnamefont
  {Dahlstr{\"o}m}}, \bibinfo {author} {\bibfnamefont {A.}~\bibnamefont
  {L'Huillier}}, \ and\ \bibinfo {author} {\bibfnamefont {A.}~\bibnamefont
  {Maquet}},\ }\bibfield  {title} {\enquote {\bibinfo {title} {{Introduction to
  attosecond delays in photoionization}},}\ } {} {\bibfield
  {journal} {\bibinfo  {journal} {J. Phys. B: At. Mol. Opt. Phys.}\ }\textbf
  {\bibinfo {volume} {45}},\ \bibinfo {pages} {183001} (\bibinfo {year}
  {2012})}\BibitemShut {NoStop}%
\bibitem [{\citenamefont {Dahlstr{\"o}m}\ \emph {et~al.}(2013)\citenamefont
  {Dahlstr{\"o}m}, \citenamefont {Gu{\'e}not}, \citenamefont {Kl{\"u}nder},
  \citenamefont {Gisselbrecht}, \citenamefont {Mauritsson}, \citenamefont
  {L'Huillier}, \citenamefont {Maquet},\ and\ \citenamefont
  {Ta{\"\i}eb}}]{dahlstrom2013a}%
  \BibitemOpen
  \bibfield  {author} {\bibinfo {author} {\bibfnamefont {J.}~\bibnamefont
  {Dahlstr{\"o}m}}, \bibinfo {author} {\bibfnamefont {D.}~\bibnamefont
  {Gu{\'e}not}}, \bibinfo {author} {\bibfnamefont {K.}~\bibnamefont
  {Kl{\"u}nder}}, \bibinfo {author} {\bibfnamefont {M.}~\bibnamefont
  {Gisselbrecht}}, \bibinfo {author} {\bibfnamefont {J.}~\bibnamefont
  {Mauritsson}}, \bibinfo {author} {\bibfnamefont {A.}~\bibnamefont
  {L'Huillier}}, \bibinfo {author} {\bibfnamefont {A.}~\bibnamefont {Maquet}},
  \ and\ \bibinfo {author} {\bibfnamefont {R.}~\bibnamefont {Ta{\"\i}eb}},\
  }\bibfield  {title} {\enquote {\bibinfo {title} {Theory of attosecond delays
  in laser-assisted photoionization},}\ }\href {\doibase
  https://doi.org/10.1016/j.chemphys.2012.01.017} {\bibfield  {journal}
  {\bibinfo  {journal} {Chemical Physics}\ }\textbf {\bibinfo {volume} {414}},\
  \bibinfo {pages} {53} (\bibinfo {year} {2013})},\ \bibinfo {note} {attosecond
  spectroscopy}\BibitemShut {NoStop}%
\bibitem [{\citenamefont {Pazourek}\ \emph {et~al.}(2015)\citenamefont
  {Pazourek}, \citenamefont {Nagele},\ and\ \citenamefont
  {Burgd\"orfer}}]{pazourek2015a}%
  \BibitemOpen
  \bibfield  {author} {\bibinfo {author} {\bibfnamefont {R.}~\bibnamefont
  {Pazourek}}, \bibinfo {author} {\bibfnamefont {S.}~\bibnamefont {Nagele}}, \
  and\ \bibinfo {author} {\bibfnamefont {J.}~\bibnamefont {Burgd\"orfer}},\
  }\bibfield  {title} {\enquote {\bibinfo {title} {Attosecond chronoscopy of
  photoemission},}\ }\href {\doibase 10.1103/RevModPhys.87.765} {\bibfield
  {journal} {\bibinfo  {journal} {Rev. Mod. Phys.}\ }\textbf {\bibinfo {volume}
  {87}},\ \bibinfo {pages} {765} (\bibinfo {year} {2015})}\BibitemShut
  {NoStop}%
\bibitem [{\citenamefont {Cattaneo}\ \emph {et~al.}(2016)\citenamefont
  {Cattaneo}, \citenamefont {Vos}, \citenamefont {Lucchini}, \citenamefont
  {Gallmann}, \citenamefont {Cirelli},\ and\ \citenamefont
  {Keller}}]{cattaneo2016a}%
  \BibitemOpen
  \bibfield  {author} {\bibinfo {author} {\bibfnamefont {L.}~\bibnamefont
  {Cattaneo}}, \bibinfo {author} {\bibfnamefont {J.}~\bibnamefont {Vos}},
  \bibinfo {author} {\bibfnamefont {M.}~\bibnamefont {Lucchini}}, \bibinfo
  {author} {\bibfnamefont {L.}~\bibnamefont {Gallmann}}, \bibinfo {author}
  {\bibfnamefont {C.}~\bibnamefont {Cirelli}}, \ and\ \bibinfo {author}
  {\bibfnamefont {U.}~\bibnamefont {Keller}},\ }\bibfield  {title} {\enquote
  {\bibinfo {title} {Comparison of attosecond streaking and rabbitt},}\
  } {} {\bibfield  {journal} {\bibinfo  {journal} {Opt. Express}\
  }\textbf {\bibinfo {volume} {24}},\ \bibinfo {pages} {29060} (\bibinfo {year}
  {2016})}\BibitemShut {NoStop}%
\bibitem [{\citenamefont {Vacher}\ \emph {et~al.}(2017)\citenamefont {Vacher},
  \citenamefont {Gaillac}, \citenamefont {Maquet}, \citenamefont {Ta{\"\i}eb},\
  and\ \citenamefont {Caillat}}]{vacher2017a}%
  \BibitemOpen
  \bibfield  {author} {\bibinfo {author} {\bibfnamefont {M.}~\bibnamefont
  {Vacher}}, \bibinfo {author} {\bibfnamefont {R.}~\bibnamefont {Gaillac}},
  \bibinfo {author} {\bibfnamefont {A.}~\bibnamefont {Maquet}}, \bibinfo
  {author} {\bibfnamefont {R.}~\bibnamefont {Ta{\"\i}eb}}, \ and\ \bibinfo
  {author} {\bibfnamefont {J.}~\bibnamefont {Caillat}},\ }\bibfield  {title}
  {\enquote {\bibinfo {title} {Transition dynamics in two-photon ionisation},}\
  }\href {\doibase 10.1088/2040-8986/aa8f56} {\bibfield  {journal} {\bibinfo
  {journal} {J. Opt.}\ }\textbf {\bibinfo {volume} {19}},\ \bibinfo {pages}
  {114011} (\bibinfo {year} {2017})}\BibitemShut {NoStop}%
\bibitem [{\citenamefont {Kheifets}(2023)}]{kheifeits2023a}%
  \BibitemOpen
  \bibfield  {author} {\bibinfo {author} {\bibfnamefont {A.~S.}\ \bibnamefont
  {Kheifets}},\ }\bibfield  {title} {\enquote {\bibinfo {title} {Wigner time
  delay in atomic photoionization},}\ } {} {\bibfield  {journal}
  {\bibinfo  {journal} {J. Phys. B: At. Mol. Opt. Phys.}\ }\textbf {\bibinfo
  {volume} {56}} (\bibinfo {year} {2023})}\BibitemShut {NoStop}%
\bibitem [{\citenamefont {Gaillac}\ \emph {et~al.}(2016)\citenamefont
  {Gaillac}, \citenamefont {Vacher}, \citenamefont {Maquet}, \citenamefont
  {Ta\"{\i}eb},\ and\ \citenamefont {Caillat}}]{gaillac2016a}%
  \BibitemOpen
  \bibfield  {author} {\bibinfo {author} {\bibfnamefont {R.}~\bibnamefont
  {Gaillac}}, \bibinfo {author} {\bibfnamefont {M.}~\bibnamefont {Vacher}},
  \bibinfo {author} {\bibfnamefont {A.}~\bibnamefont {Maquet}}, \bibinfo
  {author} {\bibfnamefont {R.}~\bibnamefont {Ta\"{\i}eb}}, \ and\ \bibinfo
  {author} {\bibfnamefont {J.}~\bibnamefont {Caillat}},\ }\bibfield  {title}
  {\enquote {\bibinfo {title} {Attosecond photoemission dynamics encoded in
  real-valued continuum wave functions},}\ }\href {\doibase
  10.1103/PhysRevA.93.013410} {\bibfield  {journal} {\bibinfo  {journal} {Phys.
  Rev. A}\ }\textbf {\bibinfo {volume} {93}},\ \bibinfo {pages} {013410}
  (\bibinfo {year} {2016})}\BibitemShut {NoStop}%
\bibitem [{\citenamefont {Saalmann}\ and\ \citenamefont
  {Rost}(2023)}]{saalmann2023a}%
  \BibitemOpen
  \bibfield  {author} {\bibinfo {author} {\bibfnamefont {U.}~\bibnamefont
  {Saalmann}}\ and\ \bibinfo {author} {\bibfnamefont {J.~M.}\ \bibnamefont
  {Rost}},\ }\bibfield  {title} {\enquote {\bibinfo {title} {{Time delays in
  anisotropic systems}},}\ } {} {\bibfield  {journal} {\bibinfo
  {journal} {{digital preprint}}\ ,\ \bibinfo {pages} {arXiv:2309.02059}}
  (\bibinfo {year} {2023})}\BibitemShut {NoStop}%
\bibitem [{\citenamefont {Kulander}\ \emph {et~al.}(1992)\citenamefont
  {Kulander}, \citenamefont {Schaffer},\ and\ \citenamefont
  {Krause}}]{kulander1992a}%
  \BibitemOpen
  \bibfield  {author} {\bibinfo {author} {\bibfnamefont {K.~C.}\ \bibnamefont
  {Kulander}}, \bibinfo {author} {\bibfnamefont {K.~J.}\ \bibnamefont
  {Schaffer}}, \ and\ \bibinfo {author} {\bibfnamefont {J.~L.}\ \bibnamefont
  {Krause}},\ }\bibfield  {title} {\enquote {\bibinfo {title} {{Time-dependent
  studies of multiphoton processes}},}\ }in\  {} {\emph {\bibinfo
  {booktitle} {{Atoms in Intense Laser Fields}}}},\ \bibinfo {editor} {edited
  by\ \bibinfo {editor} {\bibfnamefont {M.}~\bibnamefont {Gavrila}}}\ (\bibinfo
   {publisher} {{Academic Press}},\ \bibinfo {year} {1992})\ p.\ \bibinfo
  {pages} {247}\BibitemShut {NoStop}%
\bibitem [{\citenamefont {Wang}\ \emph {et~al.}(2022)\citenamefont {Wang},
  \citenamefont {Zhang}, \citenamefont {Cao}, \citenamefont {Li}, \citenamefont
  {Liu},\ and\ \citenamefont {Lu}}]{wang2022a}%
  \BibitemOpen
  \bibfield  {author} {\bibinfo {author} {\bibfnamefont {R.}~\bibnamefont
  {Wang}}, \bibinfo {author} {\bibfnamefont {Q.}~\bibnamefont {Zhang}},
  \bibinfo {author} {\bibfnamefont {C.}~\bibnamefont {Cao}}, \bibinfo {author}
  {\bibfnamefont {M.}~\bibnamefont {Li}}, \bibinfo {author} {\bibfnamefont
  {K.}~\bibnamefont {Liu}}, \ and\ \bibinfo {author} {\bibfnamefont
  {P.}~\bibnamefont {Lu}},\ }\bibfield  {title} {\enquote {\bibinfo {title}
  {Helicity dependent wigner phase shift for photoionization in a circularly
  polarized laser field},}\ } {} {\bibfield  {journal} {\bibinfo
  {journal} {J. Phys. B: At. Mol. Opt. Phys.}\ }\textbf {\bibinfo {volume}
  {55}},\ \bibinfo {pages} {115001} (\bibinfo {year} {2022})}\BibitemShut
  {NoStop}%
\bibitem [{\citenamefont {Boll}\ \emph {et~al.}(2023)\citenamefont {Boll},
  \citenamefont {Martini}, \citenamefont {Palacios},\ and\ \citenamefont
  {Foj\'on}}]{boll2023a}%
  \BibitemOpen
  \bibfield  {author} {\bibinfo {author} {\bibfnamefont {D.~I.~R.}\
  \bibnamefont {Boll}}, \bibinfo {author} {\bibfnamefont {L.}~\bibnamefont
  {Martini}}, \bibinfo {author} {\bibfnamefont {A.}~\bibnamefont {Palacios}}, \
  and\ \bibinfo {author} {\bibfnamefont {O.~A.}\ \bibnamefont {Foj\'on}},\
  }\bibfield  {title} {\enquote {\bibinfo {title} {Two-color polarization
  control of angularly resolved attosecond time delays},}\ } {}
  {\bibfield  {journal} {\bibinfo  {journal} {Phys. Rev. A}\ }\textbf {\bibinfo
  {volume} {107}},\ \bibinfo {pages} {043113} (\bibinfo {year}
  {2023})}\BibitemShut {NoStop}%
\bibitem [{\citenamefont {Kheifets}\ and\ \citenamefont
  {Xu}(2023)}]{kheifets2023b}%
  \BibitemOpen
  \bibfield  {author} {\bibinfo {author} {\bibfnamefont {A.~S.}\ \bibnamefont
  {Kheifets}}\ and\ \bibinfo {author} {\bibfnamefont {Z.}~\bibnamefont {Xu}},\
  }\bibfield  {title} {\enquote {\bibinfo {title} {Polarization control of
  rabbitt in noble gas atoms},}\ }\href {\doibase 10.1088/1361-6455/ace574}
  {\bibfield  {journal} {\bibinfo  {journal} {J. Phys. B: At. Mol. Opt. Phys.}\
  }\textbf {\bibinfo {volume} {56}},\ \bibinfo {pages} {155601} (\bibinfo
  {year} {2023})}\BibitemShut {NoStop}%
\bibitem [{\citenamefont {Han}\ \emph {et~al.}(2023)\citenamefont {Han},
  \citenamefont {Ji}, \citenamefont {Ueda},\ and\ \citenamefont
  {W\"{o}rner}}]{han2023a}%
  \BibitemOpen
  \bibfield  {author} {\bibinfo {author} {\bibfnamefont {M.}~\bibnamefont
  {Han}}, \bibinfo {author} {\bibfnamefont {J.-B.}\ \bibnamefont {Ji}},
  \bibinfo {author} {\bibfnamefont {K.}~\bibnamefont {Ueda}}, \ and\ \bibinfo
  {author} {\bibfnamefont {H.~J.}\ \bibnamefont {W\"{o}rner}},\ }\bibfield
  {title} {\enquote {\bibinfo {title} {Attosecond metrology in circular
  polarization},}\ } {} {\bibfield  {journal} {\bibinfo  {journal}
  {Optica}\ }\textbf {\bibinfo {volume} {10}},\ \bibinfo {pages} {1044}
  (\bibinfo {year} {2023})}\BibitemShut {NoStop}%
\bibitem [{\citenamefont {Eberly}(1965)}]{eberly1965a}%
  \BibitemOpen
  \bibfield  {author} {\bibinfo {author} {\bibfnamefont {J.~H.}\ \bibnamefont
  {Eberly}},\ }\bibfield  {title} {\enquote {\bibinfo {title} {{Quantum
  Scattering Theory in One Dimension}},}\ } {} {\bibfield  {journal}
  {\bibinfo  {journal} {American Journal of Physics}\ }\textbf {\bibinfo
  {volume} {33}},\ \bibinfo {pages} {771} (\bibinfo {year} {1965})}\BibitemShut
  {NoStop}%
\bibitem [{\citenamefont {Form{\'a}nek}(1976)}]{formanek1976a}%
  \BibitemOpen
  \bibfield  {author} {\bibinfo {author} {\bibfnamefont {J.}~\bibnamefont
  {Form{\'a}nek}},\ }\bibfield  {title} {\enquote {\bibinfo {title} {On phase
  shift analysis of one-dimensional scattering},}\ }\href {\doibase
  http://dx.doi.org/10.1119/1.10312} {\bibfield  {journal} {\bibinfo  {journal}
  {Am. J. Phys.}\ }\textbf {\bibinfo {volume} {44}},\ \bibinfo {pages} {778}
  (\bibinfo {year} {1976})}\BibitemShut {NoStop}%
\bibitem [{\citenamefont {Sassoli~de Bianchi}(1994)}]{sassoli1994a}%
  \BibitemOpen
  \bibfield  {author} {\bibinfo {author} {\bibfnamefont {M.}~\bibnamefont
  {Sassoli~de Bianchi}},\ }\bibfield  {title} {\enquote {\bibinfo {title}
  {{Levinson's theorem, zero-resonances, and time delay in one-dimensional
  scattering systems}},}\ }\href {\doibase http://dx.doi.org/10.1063/1.530481}
  {\bibfield  {journal} {\bibinfo  {journal} {J. Math. Phys.}\ }\textbf
  {\bibinfo {volume} {35}},\ \bibinfo {pages} {2719} (\bibinfo {year}
  {1994})}\BibitemShut {NoStop}%
\bibitem [{\citenamefont {Nogami}\ and\ \citenamefont
  {Ross}(1996)}]{nogami1996a}%
  \BibitemOpen
  \bibfield  {author} {\bibinfo {author} {\bibfnamefont {Y.}~\bibnamefont
  {Nogami}}\ and\ \bibinfo {author} {\bibfnamefont {C.~K.}\ \bibnamefont
  {Ross}},\ }\bibfield  {title} {\enquote {\bibinfo {title} {{Scattering from a
  nonsymmetric potential in one dimension as a coupled-channel problem}},}\
  }\href {\doibase http://dx.doi.org/10.1119/1.18123} {\bibfield  {journal}
  {\bibinfo  {journal} {Am. J. Phys.}\ }\textbf {\bibinfo {volume} {64}},\
  \bibinfo {pages} {923} (\bibinfo {year} {1996})}\BibitemShut {NoStop}%
\bibitem [{\citenamefont {Barlette}\ \emph {et~al.}(2001)\citenamefont
  {Barlette}, \citenamefont {Leite},\ and\ \citenamefont
  {Adhikari}}]{barlette2001a}%
  \BibitemOpen
  \bibfield  {author} {\bibinfo {author} {\bibfnamefont {V.~E.}\ \bibnamefont
  {Barlette}}, \bibinfo {author} {\bibfnamefont {M.~M.}\ \bibnamefont {Leite}},
  \ and\ \bibinfo {author} {\bibfnamefont {S.~K.}\ \bibnamefont {Adhikari}},\
  }\bibfield  {title} {\enquote {\bibinfo {title} {Integral equations of
  scattering in one dimension},}\ }\href {\doibase
  http://dx.doi.org/10.1119/1.1371011} {\bibfield  {journal} {\bibinfo
  {journal} {Am. J. Phys.}\ }\textbf {\bibinfo {volume} {69}},\ \bibinfo
  {pages} {1010} (\bibinfo {year} {2001})}\BibitemShut {NoStop}%
\bibitem [{\citenamefont {Boya}(2008)}]{boya2008a}%
  \BibitemOpen
  \bibfield  {author} {\bibinfo {author} {\bibfnamefont {L.~J.}\ \bibnamefont
  {Boya}},\ }\bibfield  {title} {\enquote {\bibinfo {title}
  {{Quantum-mechanical scattering in one dimension}},}\ }\href {\doibase
  10.1393/ncr/i2008-10030-4} {\bibfield  {journal} {\bibinfo  {journal} {La
  Rivista del Nuovo Cimento}\ }\textbf {\bibinfo {volume} {31}},\ \bibinfo
  {pages} {75} (\bibinfo {year} {2008})}\BibitemShut {NoStop}%
\bibitem [{\citenamefont {Elghazawy}\ and\ \citenamefont
  {Greene}(2023)}]{elghazawy2023a}%
  \BibitemOpen
  \bibfield  {author} {\bibinfo {author} {\bibfnamefont {K.~I.}\ \bibnamefont
  {Elghazawy}}\ and\ \bibinfo {author} {\bibfnamefont {C.~H.}\ \bibnamefont
  {Greene}},\ }\bibfield  {title} {\enquote {\bibinfo {title} {Wigner time
  delay in photoionization: a 1d model study},}\ }\href {\doibase
  10.1088/1361-6455/aceb28} {\bibfield  {journal} {\bibinfo  {journal} {J.
  Phys. B: At. Mol. Opt. Phys.}\ }\textbf {\bibinfo {volume} {56}},\ \bibinfo
  {pages} {175201} (\bibinfo {year} {2023})}\BibitemShut {NoStop}%
\bibitem [{\citenamefont {Goldstein}(1980)}]{goldstein1980a}%
  \BibitemOpen
  \bibfield  {author} {\bibinfo {author} {\bibfnamefont {H.}~\bibnamefont
  {Goldstein}},\ } {} {\emph {\bibinfo {title} {Classical
  Mechanics}}}\ (\bibinfo  {publisher} {Addison-Wesley},\ \bibinfo {year}
  {1980})\BibitemShut {NoStop}%
\end{thebibliography}
%

\end{document}